\documentclass[10pt,twocolumn]{article} 
\usepackage{simpleConference}
\usepackage{times}
\usepackage{graphicx}
\usepackage{amssymb}
\usepackage{url,hyperref}
\usepackage{caption}
\usepackage{pgfplots} 
\usepackage{cuted}

\begin{document}

\title{SR-CurvANN: Advancing 3D Surface Reconstruction through Curvature-Aware Neural Networks}

\author{
    Marina Hernández-Bautista \\
    \small Department of Computer Science \\ \small and Artificial Intelligence \\
    \small Andalusian Research Institute in \\ \small  Data Science and Computational Intelligence (DaSCI), Spain\\
    \small University of Granada, Spain \\
    \small \today \\
    \small \texttt{marinahbau@ugr.es} 
    \and
     Francisco J. Melero \\
    \small Department of Software Engineering \\
    \small Andalusian Research Institute in \\ \small  Data Science and Computational Intelligence (DaSCI), Spain\\
    \small University of Granada, Spain \\
    \small \today \\
    \small \texttt{fjmelero@ugr.es} 
}

\maketitle
\thispagestyle{empty}

\begin{center}
\centering
\makebox[\textwidth][c]{
\includegraphics[width=1.1\textwidth]{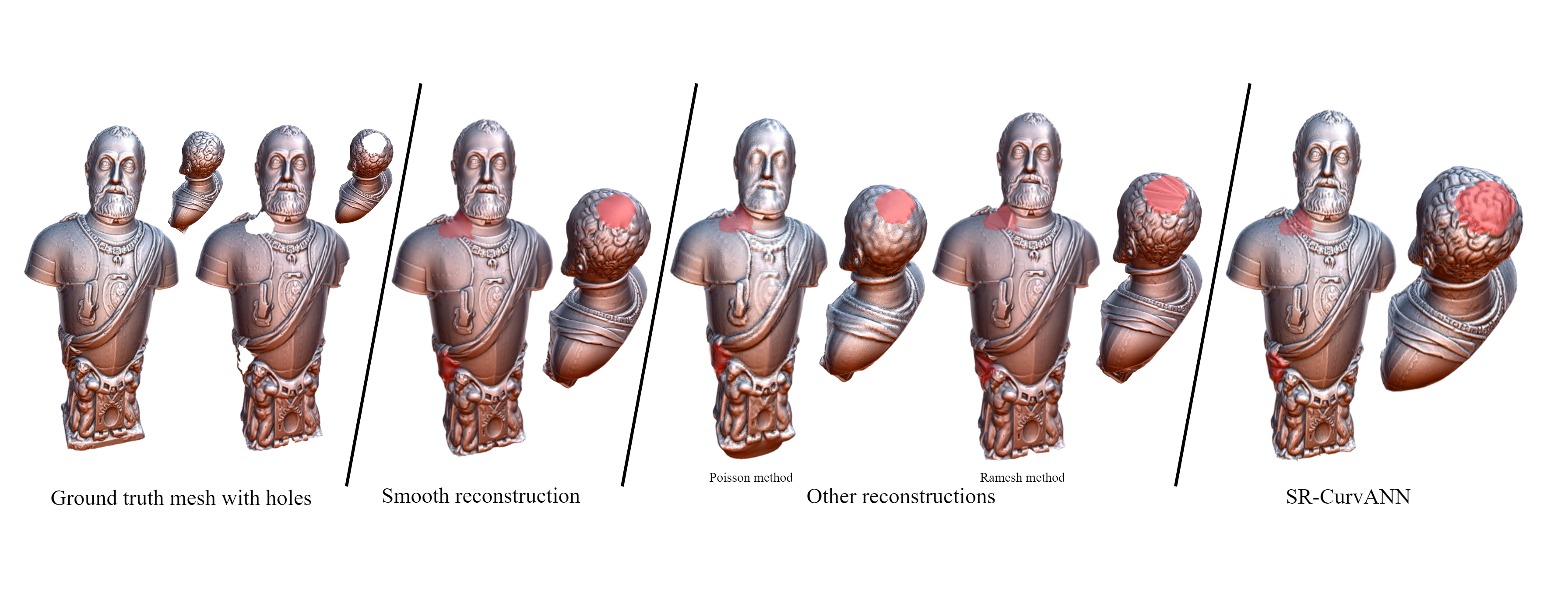}
}
\vspace{100px}
\begin{strip}
    \captionof{figure}{ \small Comparison of different shape completion techniques. The ground truth mesh with holes (leftmost) is shown alongside various reconstruction methods. Smooth reconstruction using a popular approach \cite{liepa2003filling}, Poisson method \cite{kazhdan2013screened} and Ramesh method \cite{centin2015rameshcleaner} (these methods offer the best mesh distance measures for the hole filling task) and SR-CurvANN (rightmost). The proposed method demonstrates superior fidelity in filling complex holes, particularly in regions with intricate details, as seen in the comparison with other techniques.}
\label{fig:final}
\end{strip}
\end{center}

\begin{abstract}

%The current work presents a novel methodology for completing 3D surfaces produced from 3D digitization technologies in places where there is a scarcity of meaningful geometric data. Incomplete or missing data in these three-dimensional (3D) models can lead to erroneous or flawed renderings, limiting their usefulness in a variety of applications such as visualization, geometric computation, and 3D printing. Conventional surface estimation approaches often produce implausible results, especially when dealing with complex surfaces. To address this issue, we propose a technique that incorporates neural network-based 2D inpainting to effectively reconstruct 3D surfaces. Our customized neural networks were trained on a dataset containing over 1 million curvature images. These images show the curvature of vertices as planar representations in 2D. Furthermore, we used a coarse-to-fine surface deformation technique to improve the accuracy of the reconstructed pictures and assure surface adaptability. This strategy enables the system to learn and generalize patterns from input data, resulting in the development of precise and comprehensive three-dimensional surfaces. Our methodology excels in the shape completion process, effectively filling complex holes in three-dimensional surfaces with a remarkable level of realism and precision. 

Incomplete or missing data in three-dimensional (3D) models can lead to erroneous or flawed renderings, limiting their usefulness in applications such as visualization, geometric computation, and 3D printing. Conventional surface-repair techniques often fail to infer complex geometric details in missing areas. Neural networks successfully address hole-filling tasks in 2D images using inpainting techniques. The combination of surface reconstruction algorithms, guided by the model's curvature properties and the creativity of neural networks in the inpainting processes should provide realistic results in the hole completion task. In this paper, we propose a novel method entitled SR-CurvANN (Surface Reconstruction Based on Curvature-Aware Neural Networks) that incorporates neural network-based 2D inpainting to effectively reconstruct 3D surfaces. We train the neural networks with images that represent planar representations of the curvature at vertices of hundreds of 3D models. Once the missing areas have been inferred, a coarse-to-fine surface deformation process ensures that the surface fits the reconstructed curvature image. Our proposal makes it possible to learn and generalize patterns from a wide variety of training 3D models, generating comprehensive inpainted curvature images and surfaces. Experiments conducted on 959 models with several holes have demonstrated that SR-CurvANN excels in the shape completion process, filling holes with a remarkable level of realism and precision.
\end{abstract}

\section{Introduction}

In recent years, 3D digitization has grown in popularity, affecting a wide range of fields. This technology enables detailed digital replicas of actual objects. Architectural modeling, cultural heritage preservation, medical simulations, and the development of film and video game assets all make use of scanned or photogrammetric digital models. However, mesh quality is what determines the efficacy of a 3D digital model. Uncomplete surfaces can cause render artifacts, geometric computation mistakes, and 3D printing problems. 
% The quality in question is not solely associated with precision and resolution, but also encompasses the property of watertightness.

To deal with surface imperfections, researchers use both volumetric and surface-oriented approaches. To assure hole-free models, volumetric approaches convert the mesh to an intermediate point cloud, but may degrade surface resolution and lose sensitive details \cite{Botsch2010polygon}. Surface-oriented algorithms \cite{liepa2003filling, wu2008automatic} prioritize localized modifications near the hole above global model attributes. Context-based approaches simulate the surrounding surface of the missing area \cite{harary2014context, wei2010integrated, davis2002filling}.

More recently, deep learning methods have facilitated the use of generative models to tackle surface completeness. These models mostly employ point cloud representations and ignore topological information \cite{hermoza20183d, wen2020point, wang20203d, xiao2023}. They demonstrated outstanding performance in point cloud \cite{weber2024, mirzaei2023reference, sipiran2022data} and 2D image \cite{zeng2021cr} completion tasks. Convolutional neural networks (CNNs) have also demonstrated to effectively help in repairing holes by leveraging contextual information from the surrounding neighborhood in a 2D domain \cite{gisbert2023inpainting, maggiordomo2023texture}.

%%%In recent years, deep learning methods have integrated generative models to address surface completion. These models primarily use point cloud representations, disregarding topological information \cite{hermoza20183d, wen2020point, wang20203d, xiao2023}. They have shown remarkable performance in point cloud \cite{weber2024, mirzaei2023reference, sipiran2022data} and 2D image completion tasks \cite{zeng2021cr}. Recently, deep learning inpainting techniques have been used for hole-filling in disk-homeomorphic surfaces \cite{perez2021repairing, gisbert2023inpainting}, demonstrating promising outcomes by leveraging the surrounding neighborhood as contextual information for reconstruction.

Inspired by these recent advances, we propose a novel method called SR-CurvANN (Surface Reconstruction through Curvature-Aware Neural Networks). This method integrates surface-oriented techniques with curvature-aware information and leverages the creative potential of deep learning processes to address the mesh completion task. Firstly, we generate a dataset of more than one million curvature images from an extensive collection of 3D models, and then apply random hole masks to these images. Using this image database, we train inpainting neural networks to repair holes in a concrete color palette spectrum. Once trained, these ad-hoc neural networks can infer missing parts in curvature images with real holes, thereby parametrizing and repairing the patches of the incomplete mesh. Finally, we apply an automatic curvature-guided surface reconstruction algorithm to the freshly created surface patch until it aligns accurately with the expected result. By following these steps, we recreate 3D surfaces with a remarkable level of realism and accuracy, especially in areas with intricate geometry. Therefore, the primary contributions of this work might be summarized as follows:

\begin{itemize}
\item We propose a fully automated method, named SR-CurvANN, which integrates image inpainting and virtual sculpting to reconstruct surfaces with a substantial number of polygons and intricate shapes, surpassing the limitations of existing methods that typically address simpler models.
\item We show that ad-hoc trained neural networks outperform general-purpose inpainting networks on the dataset of curvature images, leading to a more accurate and precise surface completion.
\item We have defined a process to create a vast collection of training images from authentic 3D models featuring holes.
\end{itemize}

%The remainder of the paper is organized as follows. The purpose of Section \ref{section:related} is to contextualize our work and provide the most recent and pertinent literature that motivates us in our research. Subsequently, Section \ref{section:method} illustrates the SR-CurvANN method through the dataset generation, network training, and surface reconstruction processes. We conduct a quantitative and qualitative comparison of our reconstructed surfaces with some of the most advanced algorithms in Section \ref{section:results}. Finally, in Section \ref{section:conclusions}, we discuss the results, as well as the current limitations and challenges that will inspire our future research.

\section{Related Work}
\label{section:related}
The development of robust restoration algorithms within the computer graphics community has been closely correlated with the evolution of 3D acquisition methods. The presence of holes or absent data on the acquired surface is one of the most frequently encountered issues. In the past, researchers have investigated a context-aware combination of both volumetric and surface-oriented algorithms. A new category of methods that utilize generative models for hole-filling has been introduced as a result of the recent surge in deep learning techniques. These methods predominantly rely on point cloud representations of the surface \cite{yang2018foldingnet, fan2017point}, which may result in the omission of valuable topological information. Although these methods demonstrate potential, ongoing research is working to incorporate both topological and geometric information to facilitate even more precise and detail-preserving repairs.

The utilization of 2D inpainting techniques has become an essential element in the process of restoring images within the realm of computer vision. The aim of these techniques is to reconstruct missing or damaged regions within an image by leveraging information from the neighboring intact areas. The field of CNN-based 2D inpainting has experienced significant advancements due to state-of-the-art techniques that utilize convolutional neural networks (CNN) \cite{zeng2021cr, yu2018generative}. These methods have demonstrated exceptional results and have outperformed traditional patch-based approaches. The aforementioned methods possess the ability to not only complete areas with absent information, but also thoroughly examine the surrounding context of the image, reproducing textures and patterns, and facilitating the reconstruction of intricate regions with exceptional accuracy.

As previously stated, our objective is to achieve the completion of the 3D surfaces through the reconstruction of 2D curvature images derived from the parametrization of mesh patches. The restored image generated by the algorithm provides a prediction of the shape of the surface that is missing from the polygonal model. Therefore, this section will analyze the most relevant advancements in both domains: shape completion and 2D inpainting, which are the fundamental components of our proposal.

\subsection{Shape Completion} 

A significant amount of research has been undertaken to tackle the problem of filling voids or imperfections in both basic and intricate three-dimensional objects. The classification of classical approaches can be broadly divided into three categories:

\begin{itemize}
\item Surface-based approaches. The initial set of methodologies centers around utilizing the surface data of the mesh in order to fill in areas that are incomplete or missing. The estimation of the shape of missing areas is frequently accomplished through the utilization of local or global surface fitting techniques, as discussed in previous studies \cite{liepa2003filling, attene2010lightweight, centin2015rameshcleaner, botsch2004remeshing, brunton2009filling, qiang2010hole, fang1995delaunay}. Nevertheless, these techniques may not be appropriate for larger or more intricate holes.
    \item Volumetric approaches \cite{davis2002filling, guo2006filling, park2006surface,kazhdan2013screened} represent shapes by utilizing grids or voxel grids. The estimation of missing data is accomplished through the use of extrapolation or interpolation techniques applied within the volume. The surface is then represented using either dense voxel grids \cite{dai2017shape, stutz2018learning} or octrees \cite{riegler2017octnet}, which effectively address challenges associated with memory usage and computational complexity. The aforementioned methods have demonstrated notable efficacy in addressing intricate voids and have been employed in a multitude of research investigations.

     \item Context-based approaches have been widely studied and implemented in various fields. The utilization of pre-existing information or patterns within the mesh is employed to complete the missing sections. The authors in \cite{harary2014context, vichitvejpaisal2014surface} utilize contextual relationships among various components of the shape in order to fill in the gaps in the regions that are missing. Alternatively, they also concentrate on utilizing local geometric patterns or features to deduce the missing parts. The task can be accomplished using either surface-based techniques \cite{harary2014context} or volumetric techniques \cite{sharf2004context}.
\end{itemize}

As artificial intelligence has advanced, new neural network-based hole-filling techniques have appeared. These strategies use machine learning to anticipate shape information that is lacking by using patterns that have been learned from complete data that already exists. 

The most often used architectures are based on autoencoders, which are used to acquire various prior information and generate a latent space of forms that they map to their partial inputs \cite{yuan2018pcn, chu2021unsupervised, stutz2018learning, park2019deepsdf, dai2017shape}. Utilizing variational autoencoders (VAE) for training, the methods of \cite{stutz2018learning, atzmon2020sal} have proven to be accurate when applied to 3D shapes from the ShapeNet \cite{chang2015shapenet} and KITTI \cite{geiger2012we} datasets.

A coarse-to-fine approach is implemented in both \cite{dai2017shape} and \cite{yuan2018pcn} to facilitate shape completion. In the former, a 3D shape classification network is implemented to provide additional information. However, this approach may be limited in its adaptability and is contingent upon training data. Conversely, the latter abstracts a point cloud into a feature vector, resulting in a coarse-to-fine-grained output.

GAN architectures \cite{wang20203d, zhang2021unsupervised} have also been employed for 3D shape completion, in which a generator and a discriminator collaborate to acquire the features of a 3D model, similar to their application in image generation. These methods have demonstrated potential in the field of 3D shape completion. Reinforcement learning is also implemented \cite{sarmad2019rl}.

\subsection{Image inpainting}

Inpainting is the term commonly used to describe the process of completing what is missing in a 2D image. Various strategies have been employed to resolve the inpainting problem over the years, as the input images may differ in the context of the missing part or the size of the damages. Classical methods are either relying on interpolation or diffusion of the image properties \cite{ji2020image,bertalmio2000image,bertalmio2003simultaneous} or relying on exemplar-based patch completion \cite{criminisi2004region, ding2018perceptually, liang2001real,efros2001image}. Diffusion strategies have yielded favorable outcomes, particularly when applied to vast regions. Furthermore, hybrid methods \cite{bugeau2010comprehensive} integrate elements of both diffusion and patch-based techniques to resolve intricate regions of the image.

Recent research has concentrated on the utilization of neural networks to achieve superior outcomes, which has been facilitated by the emergence of deep learning. These methods are typically categorized based on their network architecture. One notable category is Generative Adversarial Network (GAN) architectures, which utilize both a generator and a discriminator network. The problem is converted into a binary classification task by this configuration, in which the discriminator endeavors to differentiate between real and generated outputs, and the generator endeavors to generate outputs that can deceive the discriminator. The integration of a variety of other structures within both the generator and discriminator networks is facilitated by the flexibility of GAN architectures, including Convolutional Neural Networks (CNNs) \cite{zeng2021cr, yu2018generative, pathak2016context} and Transformers \cite{jiang2021transgan}.

Pure CNN architectures comprise an additional category of methodologies. These methods frequently employ variations such as U-Net \cite{ronneberger2015u} and Fully Convolutional Networks (FCN) \cite{long2015fully} for image inpainting tasks. FCNs employ convolutional layers as encoders to extract high-dimensional features and reduce noise, while deconvolutional layers function as decoders to produce the reconstructed image. Some methods, such as \cite{qin2020face, zhang2017demeshnet, sasaki2017joint}, are included in this category. In contrast to FCN, U-Net has a symmetrical architecture that includes an encoder (down-sampling) path and a decoder (up-sampling) path. Various authors have suggested various approaches that are based on U-Net, including learnable bidirectional attention maps \cite{xie2019image}, partial convolution \cite{liu2018image}, and many others \cite{mao2016image, zeng2019learning}.

Given the abundance of methodologies available in the literature, we have prioritized those that generate cutting-edge outcomes. The subsequent methods are highly pertinent for our investigation due to their superior evaluation metrics on Places2 \cite{zhou2017places}, an important image dataset.

CRFill \cite{zeng2021cr} is a two-stage method that initially employs dilated convolutions to expand receptive fields for coarse prediction. A refinement network with two encoders comprises the second stage. These encoders concentrate on the capture of fine details through a contextual attention layer and hallucinate new patches.

The Co-mod-GAN model \cite{zhao2021large} utilizes modulation techniques inspired by style transfer methods to improve its performance. The central concept underlying Co-mod-GAN is the incorporation of co-modulation, an innovative technique that greatly enhances the generator's capacity to generate a wide range of visually captivating outputs. The modulation process in Co-mod-GAN is effectively guided by incorporating both the input image features and the latent vector. The approach described strikes an optimal equilibrium between producing a wide range of outputs and maintaining high visual quality, making it especially valuable in addressing extensive image completion tasks.

Another advanced inpainting technique is the Mask-Aware Transformer (MAT) \cite{li2022mat}. This method employs a transformer architecture to effectively fill in large holes by incorporating non-local information into the attention component. The MAT system also includes a module for style manipulation, which offers a range of image completion suggestions. 

Mask-Aware Dynamic Filtering (MADF) \cite{zhu2021image}, in contrast, employs a dynamic approach to generate kernels for individual convolution windows by utilizing mask features. The system utilizes a recovery decoder and multiple refinement decoders within a coarse-to-fine encoder-decoder framework.

In recent years, there have been notable advancements in the field of 2D image inpainting, particularly with the emergence of general-purpose text-to-image platforms such as DALL·E 2 \cite{ramesh2022hierarchical} and Stable Diffusion \cite{rombach2021highresolution}, which have demonstrated impressive performance. The enhanced generalization capability of these models is achieved through the utilization of an extensive training dataset and a distinctive training process. This results in more robust and accurate inpainting outcomes, as illustrated in \cite{hernandez2023deep}.

The DALL·E 2 model \cite{ramesh2022hierarchical} is composed of an encoder and a decoder. The text label is processed by a CLIP transformer \cite{radford2021learning}  to produce a condensed vector representation known as a text embedding. The text embedding is subsequently employed to generate an image embedding using either an autoregressive or diffusive prior. Following this, a diffusion decoder is employed to produce the ultimate image.

The Latent Diffusion Models (also referred to as Stable Diffusion) serve as an open source alternative to DALL·E 2. These models, as described in \cite{rombach2021highresolution}, are conditional diffusion models that have been trained for various tasks. The particular inpainting diffusion model is grounded in the Lama architecture \cite{suvorov2021resolution}, a cutting-edge approach specifically developed for the purpose of inpainting images with extensive masks. The utilization of fast Fourier convolutions in Lama enables the incorporation of global context starting from the initial layers of the neural network. Additionally, a loss function with a high receptive field is employed to encourage the consistency of global shapes. Furthermore, the model incorporates a comprehensive mask generation process in order to exploit the model's high receptive field characteristics.

\subsection{Shape Completion using Inpainting Techniques}

Significant progress has been achieved in the domain of shape completion in recent years, with a particular focus on the utilization of 2D inpainting techniques. The objective of these techniques is to restore or fill in absent or obscured portions of an object in an image by utilizing the capabilities of deep learning and neural networks to deduce and produce credible reconstructions. This subsection examines the most pertinent and advanced methodologies that closely correspond to our research, with a specific emphasis on those that employ 2D inpainting techniques to accomplish shape completion with a high level of accuracy.

A limited number of solutions that employ a variety of various approaches have been the result of recent research on this novel methodology. Several recent methods have employed the self-prior concept in the context of 3D inpainting, where recoveries are made by directly performing inpainting over the 3D surface. This approach overcomes the disadvantages of data-driven methods by learning prior knowledge from a single damaged point on the surface. This context enables us to evaluate methods such as those proposed in \cite{hattori2022learning, hattori2024learning}, which utilize Graph Convolutional Networks for mesh denoising and shape completion tasks, respectively. Point2Mesh \cite{hanocka2020point2mesh} is another remarkable method that employs a self-prior to reconstruct the entire surface from an input point cloud without the need for pretraining. In addition, the Deep Geometric Prior method \cite{williams2019deep} reconstructs various local surface patches as a manifold atlas. A comprehensive reconstruction of the surface is generated by sampling this atlas.

Within this particular domain, there is a prevalent trend of employing NeRFs (Neural Radiance Fields) as a means of representing the three-dimensional geometry and radiance of a given scene through the utilization of neural networks. For example, \cite{weber2024nerfiller} explores the application of a 2D diffusion model in the context of general scene completion tasks. The model is utilized to generate a coherent multi-view output for the completed 3D scene. In a similar vein, \cite{poole2022dreamfusion} employs a text-to-image diffusion model to facilitate the synthesis of 3D representations from textual input.

A series of approaches that employ analogous techniques, such as geometrical procedures and 2D inpainting techniques, have been chosen to facilitate the most accurate comprehension of the problem at hand. Inpainting is employed to recover fabric filaments in \cite{gisbert2023inpainting}. Depth images are repaired by both \cite{perez2021repairing, hu20203d}. The former employs a classical inpaint approach and subsequently conducts the 2D to 3D transformation, while the latter employs partial depth images that are repaired by a conditional generative network and reproject computations of consistency distances. Lastly, \cite{maggiordomo2023texture} employs a comparable methodology to ours, integrating deep learning with geometric techniques to execute inpainting on the mesh texture.

\section{Overview}\label{section:overview}

\begin{figure*}[!ht]
    \centering
    \includegraphics[width=\linewidth ]{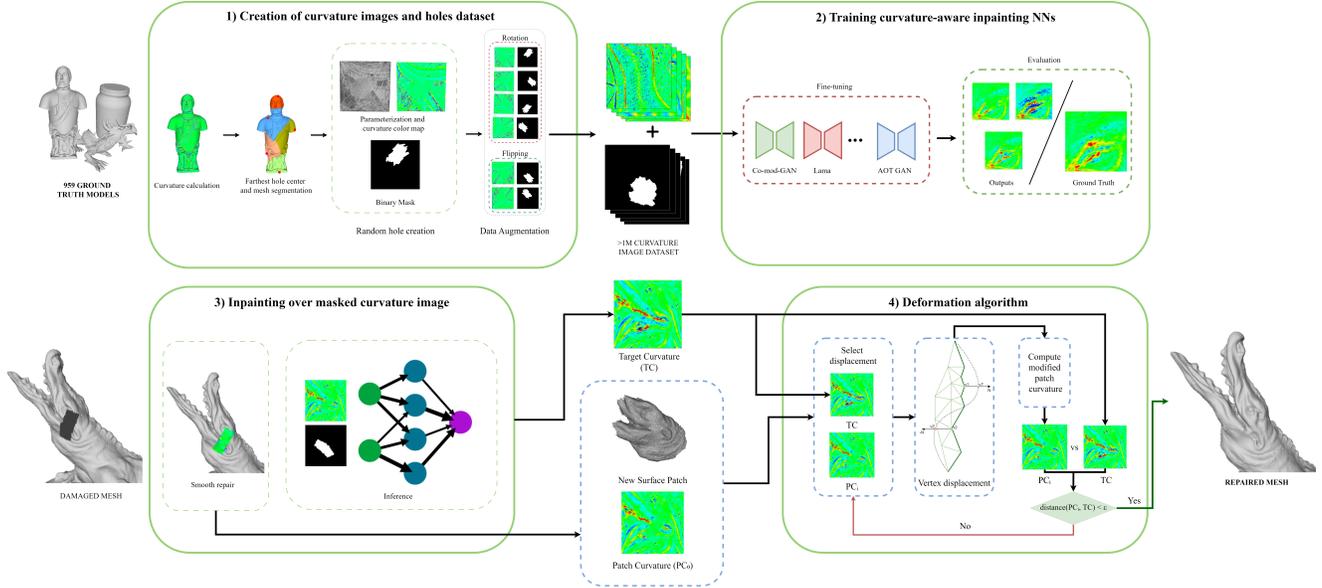}
    \caption{ \small Overview of SR-CurvANN.\textit{1) Creation of curvatures images and holes dataset}. We use a huge collection of 3D models to make a dataset of more than a million images of curvature, adding random hole masks to these images. \textit{2) Training curvature-aware inpainting NNs}. We use this database of images to teach inpainting artificial neural networks how to fix holes in a specific color palette spectrum. \textit{3) Inpainting over masked curvature image}. Once they are taught, these ad-hoc neural networks can infer missing parts of curvature images, prompting requests for them to inpaint images extracted from 3D meshes previously unused. \textit{4) Deformation Algorithm}. An automatic curvature-guided surface reconstruction algoritm is applied on the newly made surface patch until it matches the desired outcome perfectly.}
    \label{fig:overview}
\end{figure*}

This article introduces a method dubbed SR-CurvANN (Surface Reconstruction using Curvature-Aware Neural Networks) that integrates surface-based approaches and curvature-sensitive data to train deep learning algorithms for predicting the shape of the missing portion of the mesh.

As illustrated in the visual diagram of SR-CurvANN shown in Figure \ref{fig:overview}, our method may be described into four main stages:

\begin{enumerate}
    \item \textbf{Data preparation:} We extract over 1 million curvature images from a large collection of topologically accurate 3D models. The images depict the three-dimensional curvature at the vertices, shown in a two-dimensional format using a color-coded scale. In addition, each curvature image is associated with a collection of randomly generated hole masks, which enhances the diversity of the training dataset.
    
    \item \textbf{Training of neural networks:} To train the inpainting neural networks, we use the database of curvature images and their corresponding hole masks as input. Such ad-hoc trained neural networks improve the precision of the deep learning inpainting process because they are specifically built to repair images with a well-defined color range.

    \item \textbf{Inference over real damaged models:} The curvature-aware neural network receives a 2D curvature-based image from a 3D surface patch that contains a hole and is able to effectively reconstruct the missing area. The neural network infers the curvature of the missing part from the patch itself, as well as from features from other similar polygonal models, thanks to prior training.

    \item \textbf{Coarse-to-Fine surface deformation algorithm:} Upon completion of the inference phase, we apply a coarse-to-fine surface deformation algorithm on the 3D mesh. This algorithm aligns an initially simple surface patch with the expected result, guaranteeing that the reconstructed surface is comparable to the real geometry and topology of the model. As Figure \ref{fig:final} shows, this stage enables the final patch to achieve an extraordinary degree of accuracy and detail.
\end{enumerate}

\section{Method}\label{section:method}

Our SR-CurvANN method entails the creation of a curvature feature image and the subsequent application of inpainting techniques to effectively fill in the missing components. The focus of our research is on automating and improving the methodology used to address the limitations of generic inpainting networks. 

In the following, we describe in Section \ref{subsection:dataset} the selection of the 3D models that are used for the training images dataset. The entire process entails the conversion of the 3D mesh representations into 2D images, thereby guaranteeing that the resultant images accurately depict the geometrical details of the original surfaces. We will not only use these 3D meshes to produce 2D training images, but also use them as a benchmark to evaluate the effectiveness of our shape completion algorithm. Following that, in Section \ref{section:training} we explain the second stage of our method, which is implemented using different cutting-edge ad-hoc architectures for image inpainting to benchmark their accuracy. Once trained, the different inpainting approaches are subjected to evaluation, requesting them to complete 3D meshes with missing parts and comparing their performance, as depicted in Section \ref{section:test}. The automatic algorithm that conducts virtual sculpting of the mesh surface is explained in Section \ref{section:algorithm}. This method, as will be explained in Section \ref{section:results}, results in an accurate solution to the 3D model shape reconstruction problem.

\subsection{Creation of the training dataset} 
\label{subsection:dataset}

The process of creating the dataset involves multiple stages, resulting in the formation of two separate databases.

Firstly, we provide an explanation in Section \ref{subsection:dataset:3dmodels} regarding the specific criteria that were selected for the inclusion of 3D models in the dataset. These criteria ensure that the models are suitable for use as a training dataset for 2D inpainting tasks. After the selection of the 3D models, the creation of realistic holes on the surface is carried out according to the procedures described in Section \ref{subsection:dataset:holes}. The initial collection of three-dimensional models, consisting of intact meshes as well as their corresponding damaged counterparts, will be utilized for the purpose of conducting tests and obtaining an unbiased assessment of the shape completion algorithm.

As outlined in \ref{subsection:dataset:images}, the complete models undergo a secondary procedure where their 3D surfaces are partitioned into patches and subsequently parametrized to yield multiple 2D surfaces per mesh. Subsequently, artificial holes are introduced into the images, and through the implementation of traditional augmentation techniques, a dataset consisting of over one million images is obtained. This dataset is then utilized for training the inpainting neural network.

\subsubsection{Selection of 3D models} \label{subsection:dataset:3dmodels}

To ensure the success of the parametrization or reconstruction process, it is imperative that all meshes in the dataset possess the quality of being watertight, meaning they should not contain any holes. Additionally, they should be manifold, indicating that they are continuous and have no self-intersections. Furthermore, the meshes should consist of a single component with a substantial number of polygons. The aforementioned characteristics play a pivotal role, as the correct topology enables us to execute geometric operations on the mesh surface while repairing the affected areas, thereby ensuring the efficacy of our reconstruction method.

We have conducted an analysis on a comprehensive set of seven distinct 3D model datasets that are frequently referenced in academic literature \cite{chang2015shapenet, deitke2023objaverse, collins2022abo, downs2022google, koch2019abc, wu20153d, dai2017scannet}. Our investigation has revealed a number of issues, which are detailed in Table \ref{table:datasets}. The majority of the datasets fail to meet our requirement for watertightness as a result of incorrect topology. The datasets in question exhibited erroneous topology or were specifically designed for mesh segmentation purposes, resulting in the presence of disconnected faces. One frequently encountered issue is the quantity of polygons per mesh. Meshes with a limited number of faces can result in inadequate levels of detail, as models with low resolution tend to generate color images with reduced variability in patterns. Consequently, this can negatively impact the training process and result in less accurate reconstructions. Conversely, an excessive quantity of polygons will lead to a dataset that requires significant resources for both training and evaluation. In order to ensure the quality of our dataset, we implemented a filtering process to specifically select 3D models that possessed a suitable number of polygons and exhibited correct topology.

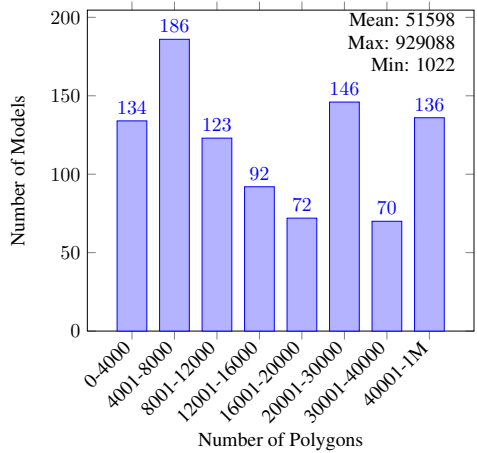
\begin{figure}[h!]
    \centering
    \begin{tikzpicture}[thick,scale=0.75]
        \begin{axis}[
            ybar,
            xlabel={Number of Polygons},
            ylabel={Number of Models},
            symbolic x coords={0-4000,4001-8000,8001-12000,12001-16000,16001-20000,20001-30000,30001-40000,40001-1M},
            xtick=data,
            x tick label style={rotate=45, anchor=east},
            bar width=15pt,
            ymin=0,
            nodes near coords,
            enlarge x limits=0.15,
            x label style={at={(axis description cs:0.5,-0.3)},anchor=south}
        ]
        \addplot coordinates {(0-4000,134) (4001-8000,186) (8001-12000,123) (12001-16000,92) (16001-20000,72) (20001-30000,146) (30001-40000,70) (40001-1M,136)};
        \node[anchor=east]  at (axis description cs:0.97,0.97) {Mean: 51598}; % Adjust position as needed
        \node[anchor=east]  at (axis description cs:0.97,0.90) {Max: 929088}; % Adjust position as needed
        \node[anchor=east]  at (axis description cs:0.97,0.83) {Min: 1022}; % Adjust position as needed
        \end{axis}
    \end{tikzpicture}
    \caption{\small The distribution of the number of polygons in the dataset of 3D models.}
    \label{fig:polygon_distribution}
\end{figure}

Upon thorough examination of the aforementioned datasets, we have successfully collated an extensive catalog consisting of 941 models out of the original 1032 models from the Google Dataset. The exclusion criteria for the remaining 91 models was based on their inability to be integrated into a single component without any modifications to the mesh topology. In order to augment our dataset with an additional set of natural figures, we integrated two models sourced from the Stanford Repository, as well as sixteen models that were digitized by the authors themselves in different projects. The final curated database consists of a collection of 959 3D manifold models that meet the essential criteria for being watertight and possessing optimal resolution. The distribution of the number of polygons across the dataset is presented in Fig. \ref{fig:polygon_distribution}.

\begin{table*}[h] 
\centering
\caption{ \small Analysis of different datasets and their challenges.}\label{table:datasets} 
\begin{tabular}{ |p{6cm}|p{10cm}|}
 \hline
  Dataset& Description and Issues\\
 \hline
 ShapeNetSem \cite{chang2015shapenet}  & 12,000 models distributed across 270 categories. The division of each face of every model in the dataset into separate components presents a notable difficulty when attempting to integrate them into a cohesive mesh for the purpose of hole-filling tasks.\\
  \hline
 Objaverse 1.0 \cite{deitke2023objaverse}& A dataset with annotations comprising more than 800,000 models. The models are divided into semantic components. The process of segmenting the meshes presents a significant difficulty when trying to merge them together for tasks like hole-filling. \\
  \hline
 Amazon Berkeley Objects \cite{collins2022abo}  & Approximately 8,000 objects belonging to 63 different categories. It includes a product catalog as well as artist-created 3D models that accurately represent various household objects, incorporating intricate geometries. The dataset presents a significant challenge in terms of resource allocation due to its substantial size of 154 GB. \\
  \hline
 Google Scanned Objects \cite{downs2022google} & The dataset consists of 1032 models that accurately depict various household items. The majority of the models are devoid of corrupt features and consist of a singular component.\\
  \hline
 ModelNet  \cite{wu20153d}& 127,000 CAD models belonging to over 600 categories. The process of exporting 3D models to CAD formats frequently leads to corruption, specifically involving non-manifold edges and vertices, as well as division into face components. The aforementioned issues make the models incompatible with our hole-filling approach, as they cannot be readily rectified.\\
 \hline
 ABC Dataset  \cite{koch2019abc} & Over 1,000,000 high quality CAD models that are accompanied by ground truth normals and curvature values. The dataset comprises explicitly parametrized curves and surfaces, which serve as a reliable reference for patch segmentation. The division of models into face components is necessitated by the use of CAD format.  \\
  \hline
 ScanNet \cite{dai2017scannet} & Dataset that comprises over 1500 annotated scans of indoor scenes. The models in this dataset are non-manifold, meaning they have irregularities in their geometry. Additionally, the dataset includes realistic holes that were generated due to scanning difficulties. However, the models are divided into semantical components, which poses challenges in accurately filling the correct holes and treating the scene as a unified component.\\
 \hline
\end{tabular}
\end{table*}

\subsubsection{Creating holes in 3D models for testing}\label{subsection:dataset:holes}

The benchmark models utilized for the shape completion algorithms necessitate the presence of one or more holes in the surface. Consequently, we were compelled to fabricate authentic holes within the meshes, as illustrated in the upper branch of Fig. \ref{fig:dataset_diagram}.

\begin{figure}[b!]
    \centering
    \includegraphics[width=\linewidth ]{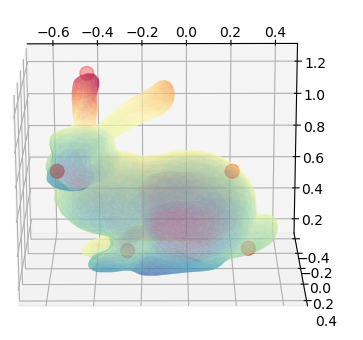}
    \caption{\small The initial 3D mesh model follows a farthest point sampling strategy, selecting the starting vertices for the perforations to ensure that each vertex is maximally distant from other vertices in the surface.}\label{fig:starting_points}
   
\end{figure}

To simulate perforations, a scripted procedure is employed for each model, whereby the mesh is initially divided into multiple evenly distributed patches. By employing this approach, we ensure that the placement of each subsequent hole center is maximally distant from the centers of the other holes. As a result, this sampling strategy enables us to target distinct regions of the surface and prevent any potential overlap between the holes.

The algorithm utilizes a farthest point sampling strategy on the surface mesh to choose five vertices in such a way that each vertex is maximally distant from the others, thereby guaranteeing a well-distributed set of starting points, as depicted in Fig. \ref{fig:starting_points}. The algorithm computes the squared Euclidean distances between each vertex and the remaining vertices, and then selects the vertices with the largest minimum distance to the vertices that have already been chosen. 

When five seed vertices are chosen, the selected vertices area is expanded iteratively, introducing unique perforations. This expansion continues until a random number of vertices (which is less than 10\% of the total number of vertices in the mesh) are selected. In the end, the process involves the elimination of individual patches made up of chosen vertices and faces, resulting in the creation of a modified representation of the 3D model that exhibits damage. 

\begin{figure}
    \centering
    \includegraphics[width=\linewidth ]{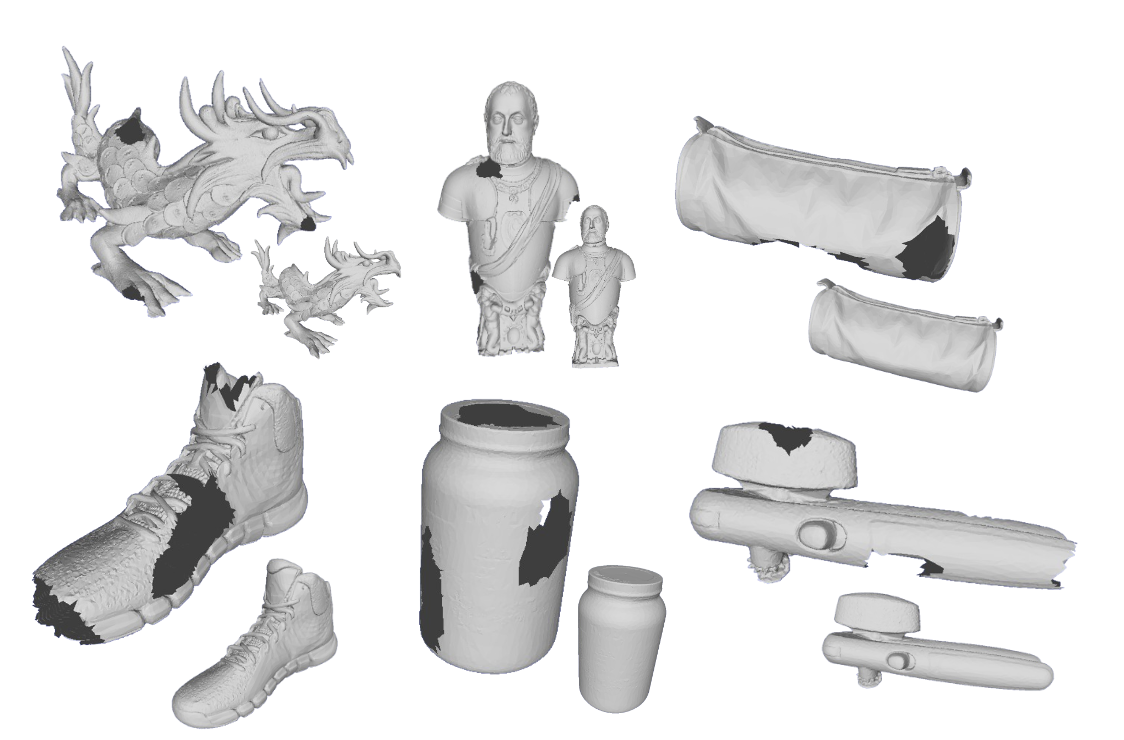}
    \caption{\small Examples of some ground truth models in the dataset and the simulated holes.}
    \label{fig:holes}
\end{figure}

By employing a systematic methodology, every model contained within this database exhibits unique and accurately simulated holes that are frequently encountered in real-life situations, as illustrated in the examples shown in Fig. \ref{fig:holes}. Currently, the database contains ground truth models that are complete and do not have any holes in them, as well as incomplete models that have organic deficiencies introduced through our perforation simulation method.

\subsubsection{Curvature images for training}\label{subsection:dataset:images}

\begin{figure*}
    \centering
    \includegraphics[width=\linewidth]{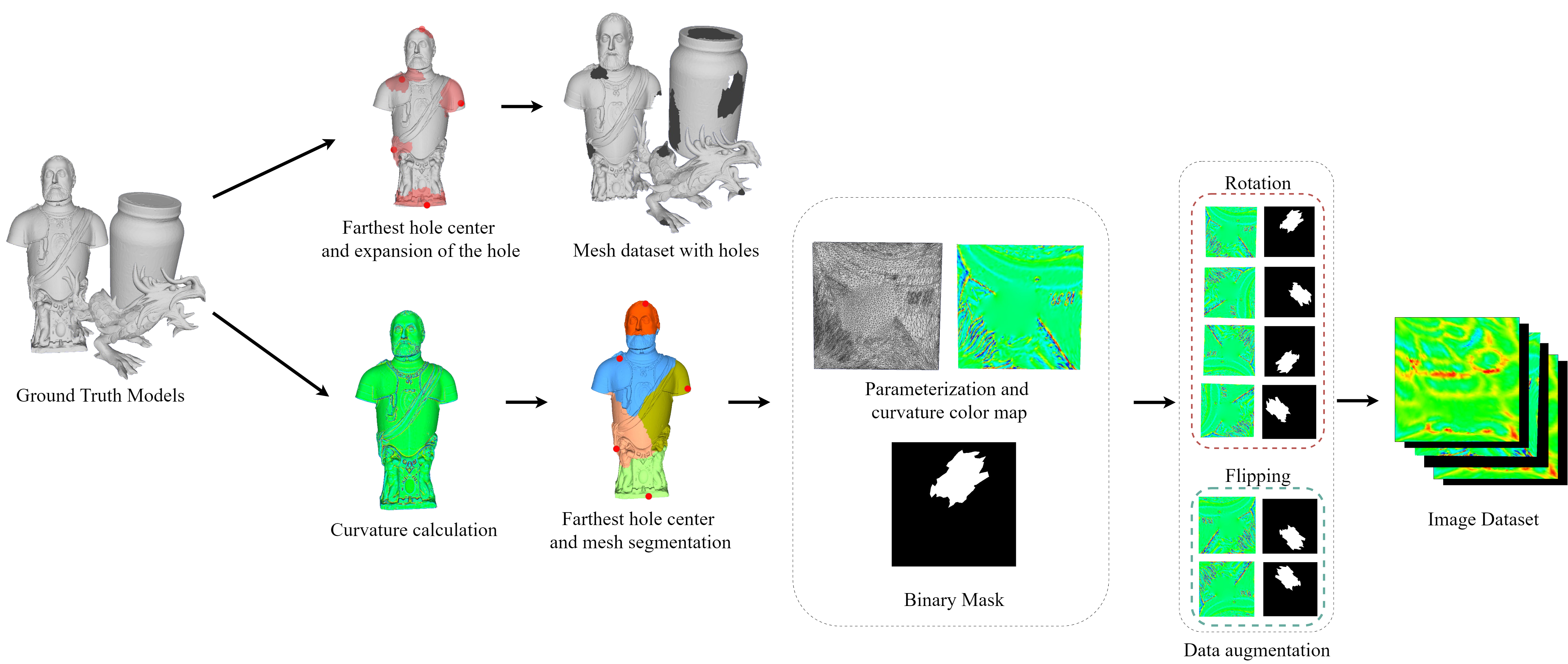}
    \caption{\small The process of creating a dataset. Uncomplete meshes are generated based on the ground truth models by utilizing a set of seed vertices and expanding the hole from these vertices (upper branch). Additionally, the dataset of images containing curvature color patches and binary masks for holes (lower branch) is generated from the ground truth models through a distinct process.}
    \label{fig:dataset_diagram}
\end{figure*}

%\emph{Parametrización, colores usados, y datos finales estadísticos... }
The ground truth models are utilized to create a dataset of images, which will subsequently be employed for training the neural network. The process is visually illustrated in the lower branch of Fig. \ref{fig:dataset_diagram}.

The central concept involves the representation of the models' surfaces as a collection of two-dimensional images. This is achieved by sampling the mean curvature values and parametrizing distinct patches for each mesh. The concept of mean curvature is used to quantify the curvature of a surface, allowing for a deeper understanding of its geometric characteristics. The mean curvature at any given point on a surface can be defined as the arithmetic mean of its two principal curvatures, which correspond to the maximum and minimum curvatures in the principal directions passing through that particular point. After obtaining the normalized mean curvature for each vertex, we proceed to assign these values to RGB tuples using a color rainbow map. In this map, the color blue corresponds to low values (indicating convexity), green represents medium values (indicating flatness), and red signifies high values (indicating concavity).

In order to segment the model and acquire the patches for parametrization, two approaches are employed. In the initial step, a predetermined quantity of seed points (2, 5, and 10) is chosen, employing a similar methodology as that used for generating the holes, in order to guarantee an equitable dispersion of the patches throughout the mesh. By strategically placing the initial points at maximum distance from one another, we are able to generate patches of uniform size that encompass the entire surface. Following the aforementioned subdivisions, a total of 17 patches are obtained per mesh.

In addition, the intact mesh is also divided into distinct significant patches, using the methodology described in \cite{shapira2008consistent}. This resulted in a variable number of patches, as it is contingent upon the shape of the 3D model. 

As proposed in \cite{hernandez2023deep}, the approach involves parameterizing the patches of the 3D models to generate 2D images that capture a partial representation of the mesh surface. Indeed, the images generated in this manner can be considered as the ground truth images, as they accurately depict the curvature of the submesh surface. 

The subsequent stage involves the generation of visual impairments, specifically the introduction of voids within the images. It is crucial to observe that the method employed does not involve utilizing the previously mentioned holes, but rather involves the creation of novel and distinct holes. By employing this approach, we ensure that the inpainting network is not trained using the voids generated in the 3D meshes. Utilizing the 3D model dataset for evaluation purposes will be facilitated by this.

\begin{figure}[h]
    \centering
    \includegraphics[width=\linewidth ]{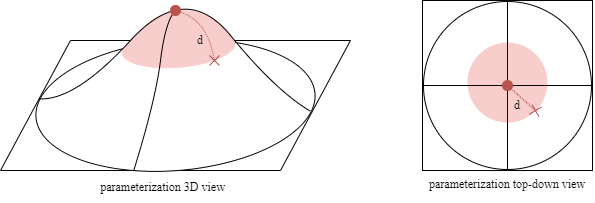}
    \caption{ \small Two perspectives on the process of arbitrarily selecting the hole center: in order to fill the holes, we randomly select an angle from 0-360 degrees and a distance that is modeled by a normal distribution. This increases the likelihood that the holes' centers are located near the image's center. The cavity is generated by expanding to the surrounding vertices from this center. This process facilitates the generation of organic random holes and guarantees that the holes generated in the meshes for inference are distinct from those in the training images.}
    \label{fig:normal_dist}
\end{figure}

The curvature image patch is not directly damaged; instead, a separate image is created to represent the hole mask. The creation of each hole mask follows a similar approach to that used on the 3D mesh. A seed vertex is randomly chosen within a central region of the image. The selection process is conducted with a certain level of randomness. The entryway is more likely to be positioned closer to the center of the resulting image. The process involves the selection of a random angle and a distance that follows a normal distribution, as illustrated in Fig. \ref{fig:normal_dist}.

Subsequently, the selection is systematically expanded across the surface topology until it encompasses a proportion exceeding 10 percent of the vertices within the specific concrete patch. The identified faces and vertices are eliminated from the parameterization and from the two-dimensional image.

The image that has been damaged is subsequently converted into a binary mask using black and white colors. In this mask, white regions indicate the absence of the surface, or the presence of a hole, while black regions indicate the remaining surface.

\begin{figure*}[!h]
    \centering
    \includegraphics[width=.8\linewidth ]{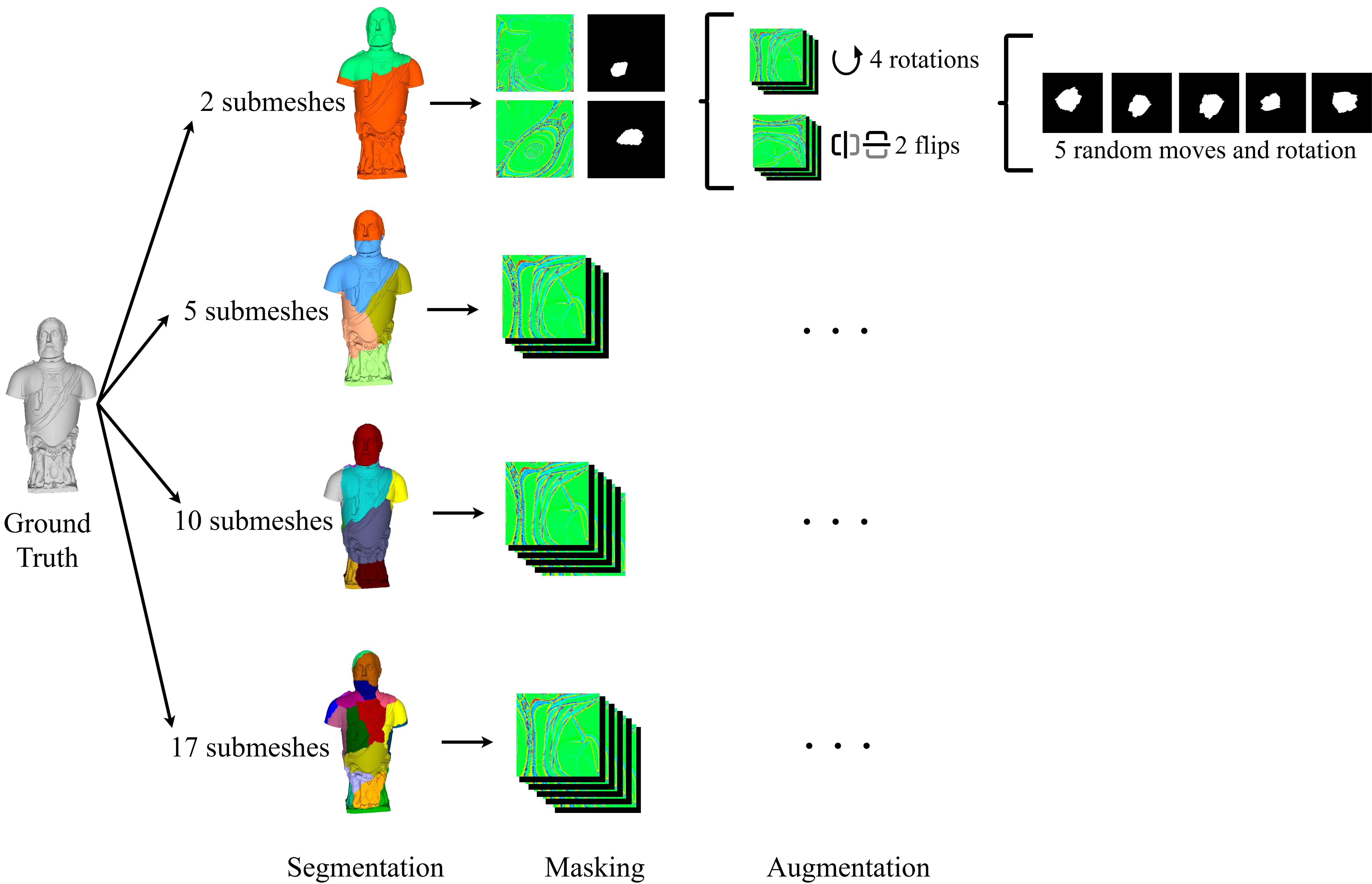}
    \caption{\small The process of generating images. The mesh is divided into distinct sections. A curvature color map and a binary mask are created for each segmentation. Subsequently, each image undergoes augmentation through the execution of four rotations and two flips. Each mask is further enhanced by executing five distinct movements and rotations of the entire area.}
    \label{fig:augmentation}
\end{figure*}

The accuracy of the trained neural network will increase as the number of images in the dataset increases. Given the limited quantity of 3D models available, which is approximately 1,000, it is necessary to perform various operations to augment the final image dataset. This is a common practice in computer vision methodologies. The paired images, consisting of curvature and binary masks, undergo rotations of 90, 180, and 270 degrees, as well as vertical and horizontal flips. In addition, each mask is modified with random displacements and rotations to the white area. This results in the creation of five additional holes per patch, as illustrated in Fig. \ref{fig:augmentation}.

To summarize, the process of generating images begins with 959 models from our mesh dataset. Each model is segmented into 2, 5, 10, and, on average, 20 significant submeshes. The dataset consists of over 35,000 images, which are enhanced by applying four rotations and two flips. This augmentation process results in a total of more than 212,000 images, each associated with a hole mask. The masks are ultimately enhanced through the process of displacing and rotating the hole five times, with each rotation being done to a random degree and direction.

The final image dataset consists of a total of 2,029,660 images, specifically 1,014,830 images of curvature patches and their corresponding 1,014,830 binary hole masks.

In order to conduct the training and validation of the algorithms, the dataset has been partitioned into separate subsets for training and testing purposes, as illustrated in Table \ref{table:dataset}.

\begin{table}[h]
\centering
\caption{\small Image Dataset distribution.} \label{table:dataset}
\begin{tabular}{llr}
\hline
\textbf{Name} & \textbf{Size} & \textbf{Images} \\
\hline
Training Set & 85\% & 1729160  \\
Training Images & 90\% training & 156910  \\
Validation Images & 10\% training & 160000  \\
Visual Test Images & & 500 \\
\hline
Test Set & 15\% & 300000 \\
\hline
Total Images &  & 1014830  \\
Total Masks &  & 1014830  \\
\hline
\end{tabular}
\end{table}

\subsection{Tranining of Neural Networks} \label{section:training}

The image dataset will function as a standardized source for the training of a neural network that is specifically designed to identify curvature color patterns. In comparison to general-purpose networks such as Stable Diffusion \cite{rombach2021highresolution}, this specialization is anticipated to improve inpainting outcomes. Co-mod-GAN \cite{zhao2021large}, Lama \cite{suvorov2021resolution}, CRFill \cite{zeng2021cr}, AOT \cite{zeng2022aggregated}, and TFill \cite{zheng2022bridging} are five distinct inpainting approaches that we have refined in order to accomplish this ad-hoc training. Their capacity to train with binary masks that indicate missing pixels in the images, the public availability of their code, and their ability to generalize to our specialized color maps were all factors that were selected. Their primary characteristics include:

\begin{itemize}
    \item Co-mod-GAN is an adversarial architecture that uses co-modulation to inpaint large missing regions.
    \item Lama architecture uses Fast Fourier Convolutions to achieve better performance.
    \item CRFill introduces contextual reconstruction loss (CR loss), which promotes the plausibility of the output generated even when reconstructed using contextual information. 
    \item TFill deploys a transformer to directly capture long-range dependence.
    \item The AOT GAN architecture aggregates contextual transformations from a variety of receptive fields, enabling the capture of both informative distant image contexts and complex patterns of interest.
\end{itemize}

\begin{figure}[h]
    \centering
    \includegraphics[width=\linewidth]{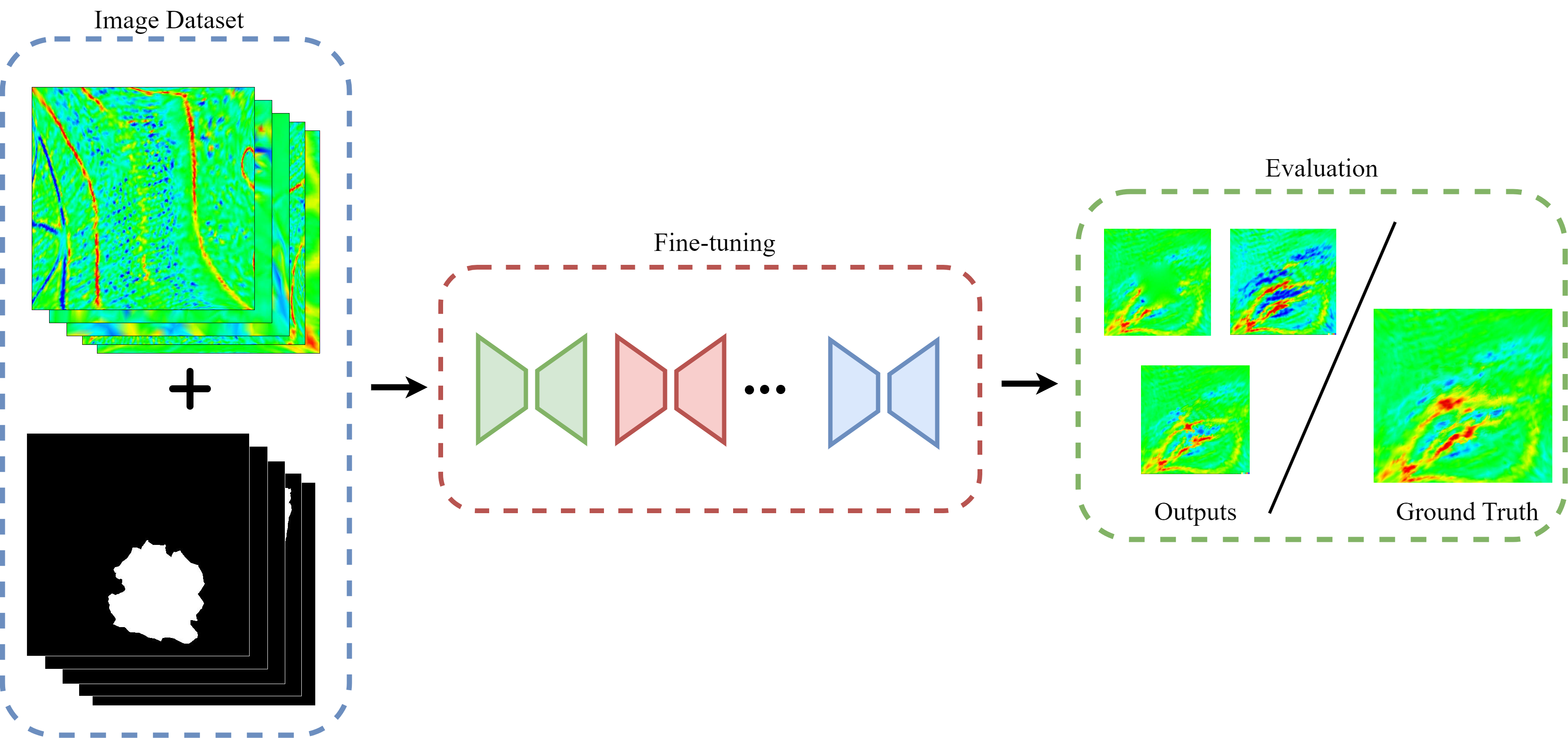}
    \caption{\small Scheme of SR-CurvANN training process. Every approach selected is fine-tuned using our image dataset and then the results are evaluated by comparing with the ground truth images. }
    \label{fig:finetuning}
\end{figure}

The fine-tuning process requires retraining each model, which has already been trained on a large dataset of general images. We use our smaller, domain-specific dataset to adapt the network specifically for the task of completing color curvature maps. This process utilizes the pre-learned features and representations from the initial training. In the previous section, we talked about how the approaches used for inpainting tasks are highly advanced, especially when it comes to the Places2 dataset \cite{zhou2017places}. The process is shown in Fig. \ref{fig:finetuning}.

\subsection{Inference over curvature images}\label{section:test}

In order to assess the effectiveness of our fine-tuning, we have chosen a range of metrics that allow us to objectively measure the quality of the predicted images. These metrics are based on widely accepted standards for evaluating image inpainting similarity:

\begin{itemize}
    \item FID (Frechet inception Distance) metric \cite{yu2021frechet} measures the distance between feature representations of real and generated images using an inception network. Lower indicates better image quality.
    \item LPIPS (Learned Perceptual Image Patch Similarity) \cite{zhang2018unreasonable} uses deep learning to calculate the perceptual difference between images. 
    \item PSNR (Peak Signal-to-Noise Ratio) \cite{hore2010image} quantifies level of noise or distortion in the reconstructed image. 
    \item SSIM (Structural Similarity Index) \cite{hore2010image} indicates image quality degradation, where higher SSIM value means higher image quality and similarity to the ground truth.  
    \item The L1 loss is a metric that measures the difference between images on a pixel level. A lower L1 loss indicates higher accuracy of the network.
    \item As in \cite{zhao2021large}, we also added P-IDS and U-IDS metrics to offer a more comprehensive comparison. These metrics align with user preferences regarding subtle differences in image quality.
\end{itemize}

\begin{figure*}[h!]
    \centering
    \includegraphics[width=\linewidth]{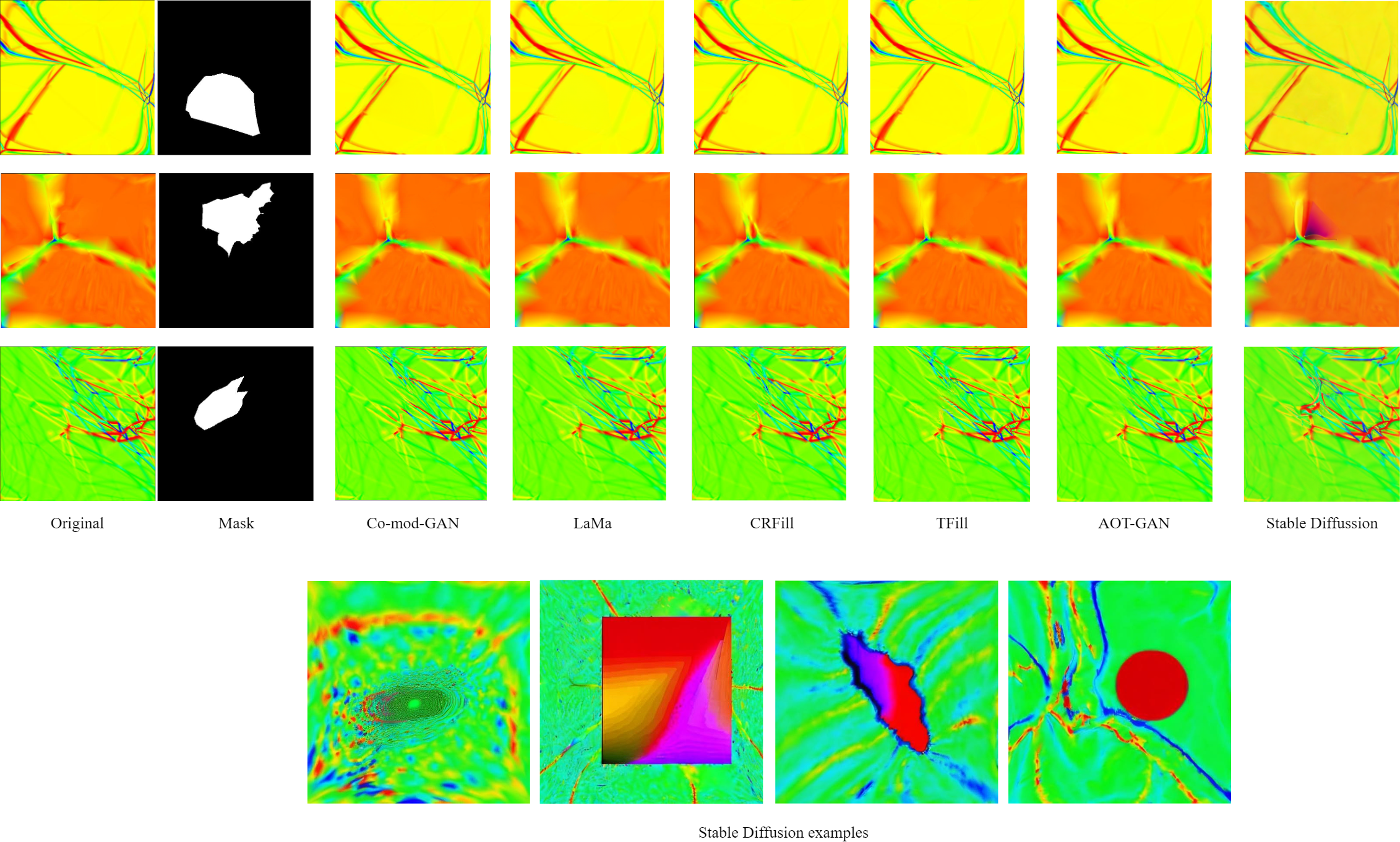}
    \caption{\small Comparison of inference results with different inpainting methods. The last row indicates some examples of unreliable inference from Stable Diffusion. }
    
    \label{fig:qual-images}
\end{figure*}

The performance of the ad-hoc trained algorithms on the curvature images is displayed in Table \ref{table:inpainting_cuantitative_table}. The metrics have been calculated both before and after performing fine-tuning. The evaluation metrics for the fine-tuned methods demonstrate their superior performance in restoring missing portions of a curvature map image compared to the generic Stable Diffusion method. In the inpainting task, Co-mod-GAN stands out as the top method, outperforming others in metrics such as FID, P-IDS, U-IDS, PSNR, and SSIM. Lama also performs well in this regard. Although Stable Diffusion produces satisfactory qualitative results, its use of general-purpose training data sometimes leads to outputs that differ greatly from the original image, as shown in Fig. \ref{fig:qual-images}.

\subsection{Coarse-to-Fine Surface Deformation Algorithm} \label{section:algorithm}

We have devised a fully automated, optimized mesh deformation process that modifies the surface to follow the predicted curvature patch, based on the work of \cite{hernandez2023deep}. This process effectively repairs the model by updating the surface with new curvature values, taking into consideration the curvature values of the neighborhood vertices of the hole and the image information from the inpainting process.

The initial step in our coarse-to-fine reparation procedure is the normalization of the mesh. Then, we employ a conventional method \cite{liepa2003filling} to fill the holes. This method triangulates the holes and refines the resulting triangulation to align with the density of triangles in the surrounding area. As was expected, the level of detail in this coarsely reconstructed region is exceedingly low.

Next, the mesh is segmented into local patches around every hole to preserve contextual information. The curvature values are used to color the per-vertex of these patches, which are parameterized to produce planar representations of the surface. The hole mask image is the result of the newly generated vertices in the previous phase.

At this stage, the neural networks must be provided with the curvature bidimensional image and its corresponding hole mask in order to execute the inpainting procedure and generate a new curvature patch image.

\begin{figure*}
    \centering
    \includegraphics[width=\linewidth]{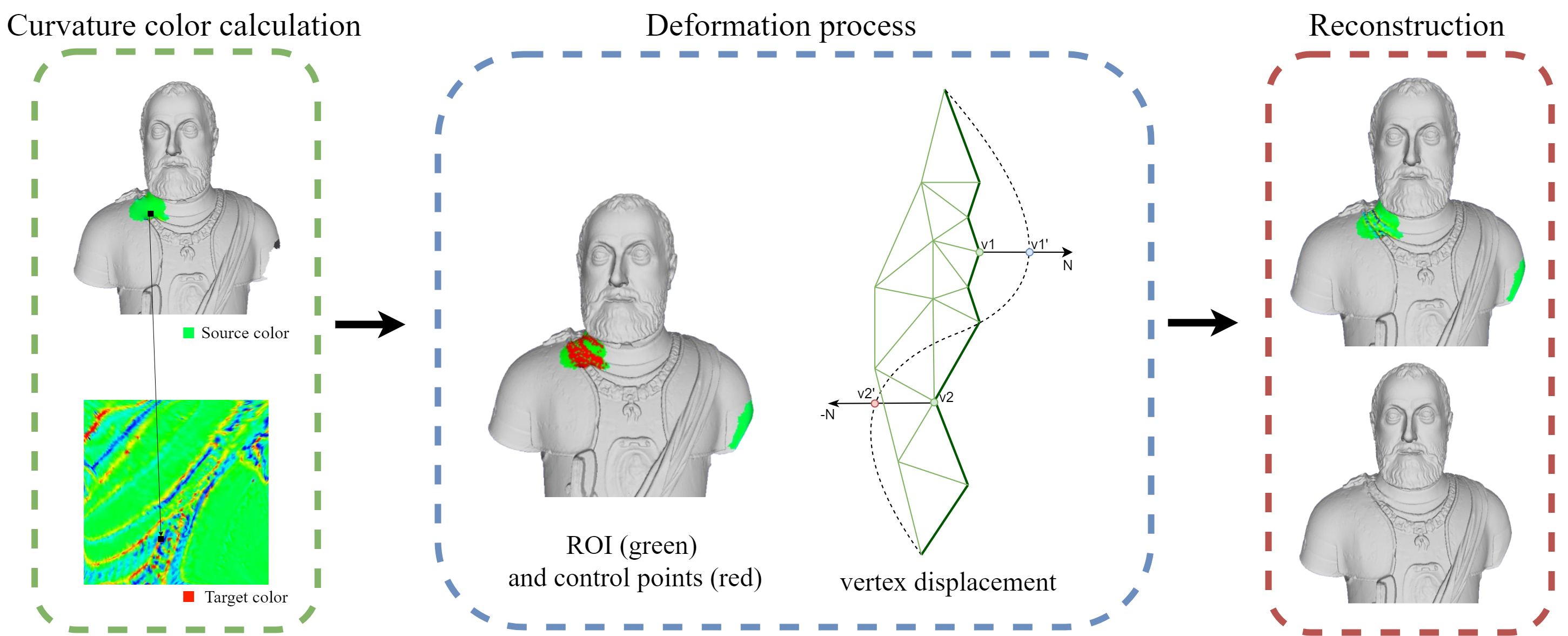}
    \caption{\small SR-CurvANN Deformation process. The curvature color calculation zone outlines the process of identifying color variations between the area where the coarse hole reconstruction takes place and the corresponding curvature inpainted image. The deformation process zone encompasses the two main aspects of deformation: the ROI, which is the coarse reconstructed patch, and a set of control points that exhibit noticeable color variations. The vertices are shifted based on the vertex displacement scheme, as the deformation requires moving them in different directions relative to the normal. Finally, we present the reconstruction following the deformation process, showcasing the updated curvature colors and the final reconstructed surface.}
    
    \label{fig:deformation-scheme}
\end{figure*}

Lastly, the surface is refined through an iterative deformation process, as illustrated in Fig. \ref{fig:deformation-scheme}. At each iteration, the new curvature image is calculated and the vertices are displaced along their normal vector. This process attempts to adjust the position of vertices by analyzing the color differences between the vertex in the curvature image of the patch and its corresponding pixel in the inferred inpainted curvature image. The direction of the vertex displacement is determined by the color distance between the two pixels. For instance, the vertices in the coarse repaired patch that correspond to green pixels (smooth curvature or flat surface) are repositioned in the direction of the normal vector if the inpainted pixel is red (convex surface) and in the opposite direction to the normal vector if the inpainted pixel indicates a concave surface area.

\begin{table*}
\centering
\caption{ \small Distance measures of different approaches across the image dataset.}\label{table:inpainting_cuantitative_table}
\begin{tabular}{ |p{3.5cm}|p{1.5cm}|p{1.5cm}|p{1.5cm}|p{1.5cm}|p{1.5cm}|p{2cm}|p{1.5cm}|  }
 \hline
  Model&FID ↓&P-IDS ↑&U-IDS ↑&LPIPS ↓&PSNR ↑&SSIM ↑ (-1,1)&L1 ↓\\
  \hline
  Co-mod-gan (original)  & 14.0258 & 0.0256 & 0.1213 & 0.0653 &inf & 0.9446 & 0.0125	\\
  Lama (original) & 3.5749 & 0.0818 & 0.255 & 0.0405 & inf & 0.9461 & 0.0111\\
 CRFill  (original) & 27.1957 & 0.0115 & 0.0757 & 0.1035 & 26.9102 & 0.9195 & 0.0169\\
 TFill (original)& 5.7126 &  0.0724 & 0.2271 & 0.0518 & 30.0261 & 0.9467 & 0.0124\\
 AOT (original) & 9.8883 & 0.0436 & 0.1953 & 0.0652 & 29.0238 & 0.9322 & 0.016\\
 Stable Diffusion (w/o fine-tuning) & 25.7099 & 0.0074 & 0.0777 & 0.1618 & 17.1572 & 0.628 & 0.1036\\
 \hline
 Co-mod-gan (fine-tuned)  & \textbf{0.8332} & \textbf{0.2816} & \textbf{0.4147} & 0.034 &\textbf{inf} & \textbf{0.9677} &	0.0086\\
 Lama (fine-tuned) & 1.109 & 0.1657 &	0.3653 & 0.0294	& inf & 0.9638 & 0.0080\\
 CRFill  (fine-tuned) & 2.345 & 0.175 & 0.3541 & 0.0398 & 32.333 & 0.9579 & 0.0096\\
 TFill (fine-tuned)& 2.5994 & 0.1423 & 0.3119 & 0.037 & 32.0994 & 0.9588 &0.0092\\
 AOT (fine-tuned) & 1.7568 & 0.1678 & 0.1678 & \textbf{0.0286} & 34.0912 & 0.9639 & \textbf{0.0079}\\
 \hline
\end{tabular}
\end{table*}

In order to preserve the original vertices of the mesh, we designate a region of interest (ROI) for the deformation algorithm as a portion of the reconstructed surface. We identify the vertices within the ROI that exhibit the most substantial color discrepancies between the curvature color and the corresponding pixel color of the inpainted image. Then, we identify the vertices whose color difference exceeds a threshold of 30 units in any color channel, such as (0, 254, 228) vs. (28, 250, 5). These vertices serve as control points for the deformation process, which is conducted using an elastic deformation approach that is based on \cite{chao2010simple}. The direction and magnitude of displacement for the entire ROI area are determined by the distance of the control locations and the sign of the color difference, as previously mentioned. Vertex positions are updated in response to applied forces, resulting in deformations that appear realistic.

Upon the algorithm's convergence, the ROI boundaries are subsequently smoothed to ensure a fluid transition between the original and new portions of the surface.

\begin{figure}
    \centering
    \includegraphics[width=\linewidth]{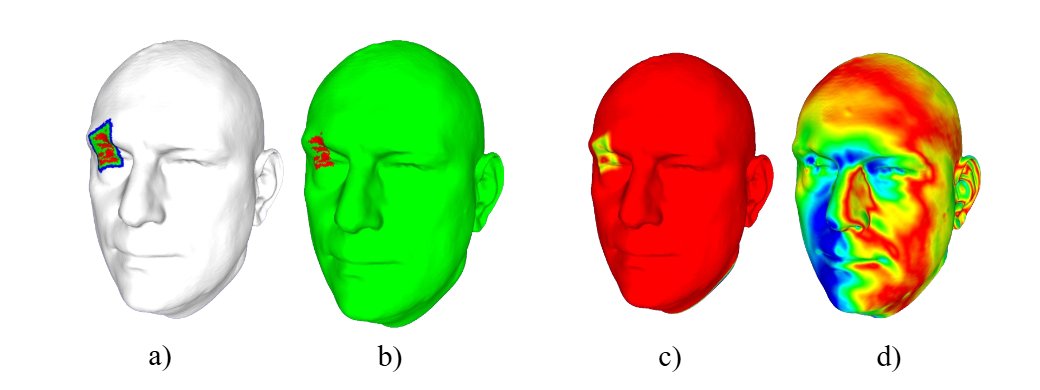}
    \caption{ \small a) ROI (green) defined as the entire patch. The control vertices are colored in red, while the smoothed boundaries are colored in blue. 
    b) ROI (green) defined as the entire mesh. c) Repaired mesh by using reduced ROI d) Repaired mesh by using ROI as the entire model. c) and d) show the colored distance of the repaired mesh to the ground truth model, with red indicating the minimum distance. }
    \label{fig:ablation}
\end{figure}

We conducted an ablation study in Fig. \ref{fig:ablation} to validate our hole-centric reconstruction method and emphasize the significance of the ROI. When the deformation is not constrained to the ROI and the elastic deformation takes into account the entire mesh, we can observe visually similar outcomes. On the other hand, the distance metrics represented by colors in each vertex show significant color variations throughout the mesh when considering a large region of interest (Fig. \ref{fig:ablation}.d). This indicates that vertices outside of the new patch have been moved (shown in yellow and blue) when they should have remained unchanged. When applying the algorithm with a modified ROI (Fig. \ref{fig:ablation}.c), all vertices remain unchanged (red) while only the hole area is repaired.

\section{Experiments and evaluation}\label{section:results}

SR-CurvANN has been assessed quantitatively and qualitatively through a series of experiments. In order to guarantee impartial and equitable comparisons and measurements, we offer a concise summary of the experimental conditions in Section \ref{subsec:conditions}. Consequently, in Section \ref{subsec:qualitative}, we examine the qualitative experiments and present the corresponding results. Lastly, we investigate the quantitative experiments in Section \ref{subsec:quantitative} to illustrate the quality of the reconstructed meshes and the performance of our approach.

\subsection{Experiment conditions}\label{subsec:conditions}
We opted to establish a restricted inference time of 9 minutes for running our inpainting neural networks. This choice allows for equitable comparisons between our approach and other cutting-edge methods. The hole completion of our complete dataset, which includes 959 models, was exceuted in approximately one week, as each mesh correction operation needs a maximum of nine minutes.

We conducted our experiments by applying the two inpainting methods that produced the most favorable visual loss metrics to the curvature images: Lama and Co-Mod-GAN (refer to Table \ref{table:inpainting_cuantitative_table}). The performance of our approach is compared to classical techniques that have software available for comparison purposes, as well as some deep learning-based methods for hole completion, in our analysis. In particular, we compared our findings to well-known benchmarks in the field, including MeshFix \cite{attene2010lightweight}, Ramesh \cite{centin2015rameshcleaner}, and Screened Poisson \cite{kazhdan2013screened} for classical methods. For deep learning-based methods, we introduced SAL \cite{atzmon2020sal} and Point2Mesh \cite{hanocka2020point2mesh}, two approaches that were developed to perform shape completion tasks. 

We have conducted a comprehensive qualitative and quantitative evaluation of the results after executing our algorithm on the incomplete meshes that were reserved for testing.  The visual appearance and similarity of details of the new patches are the primary focus of the qualitative evaluation, while the distances between the repaired mesh and the ground truth mesh are computed for the quantitative evaluation.

\subsection{Qualitative Results}\label{subsec:qualitative}

For qualitative assessment, we have selected four baseline models that have distinct features. These four models showcase a variety of organic forms and both artificial and real-world scenarios. They include a real human face, the classical Armadillo mesh from the Stanford repository, a scanned baroque 3D sculpture, and a Christmas bear figure. Our models have been through a rigorous process to create authentic-looking holes on their surfaces. We then repaired them using various methods and conducted experiments to ensure consistent results. Deep learning-based models often demand substantial computational time and graphical resources in order to produce satisfactory inference results.

\begin{figure*}[h]
    \centering
    \includegraphics[width=\linewidth]{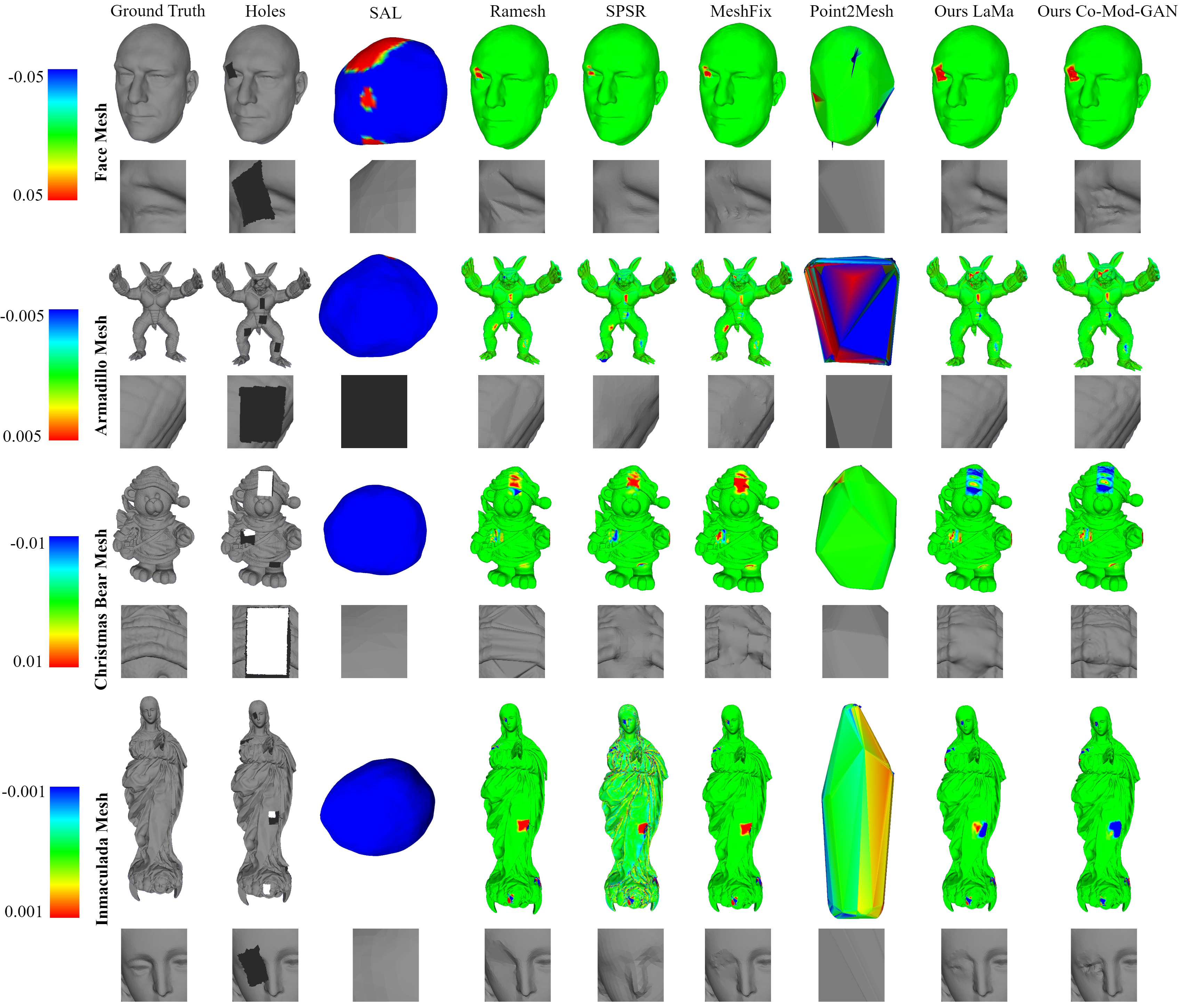}
    \caption{\small Qualitative comparison of inference through different models.  The signed distance to the ground truth mesh is indicated by the color of the vertices. Additionally, we provide a zoomed-in view of one of the holes to better examine the new patch's intricate details.}
    \label{fig:qual-ev}
\end{figure*}

Fig. \ref{fig:qual-ev} displays the results. In order to offer a more comprehensive visual analysis, we have implemented a color scheme that corresponds to the signed distance value of each mesh in relation to the ground truth mesh. It is important to note that our evaluation focuses solely on the final shape. Although the majority of the approaches yield similar results (excluding SAL and Point2Mesh), a more detailed examination of the repaired regions reveals that our approach accurately simulates more lifelike surfaces. The Virgin Mary sculpture showcases a distinct approach to reconstructing the eye, focusing on accurately recovering the original shape rather than simply extending local geometry around the hole.

\begin{figure*}[!h]
    \centering
    \includegraphics[width=0.8\linewidth]{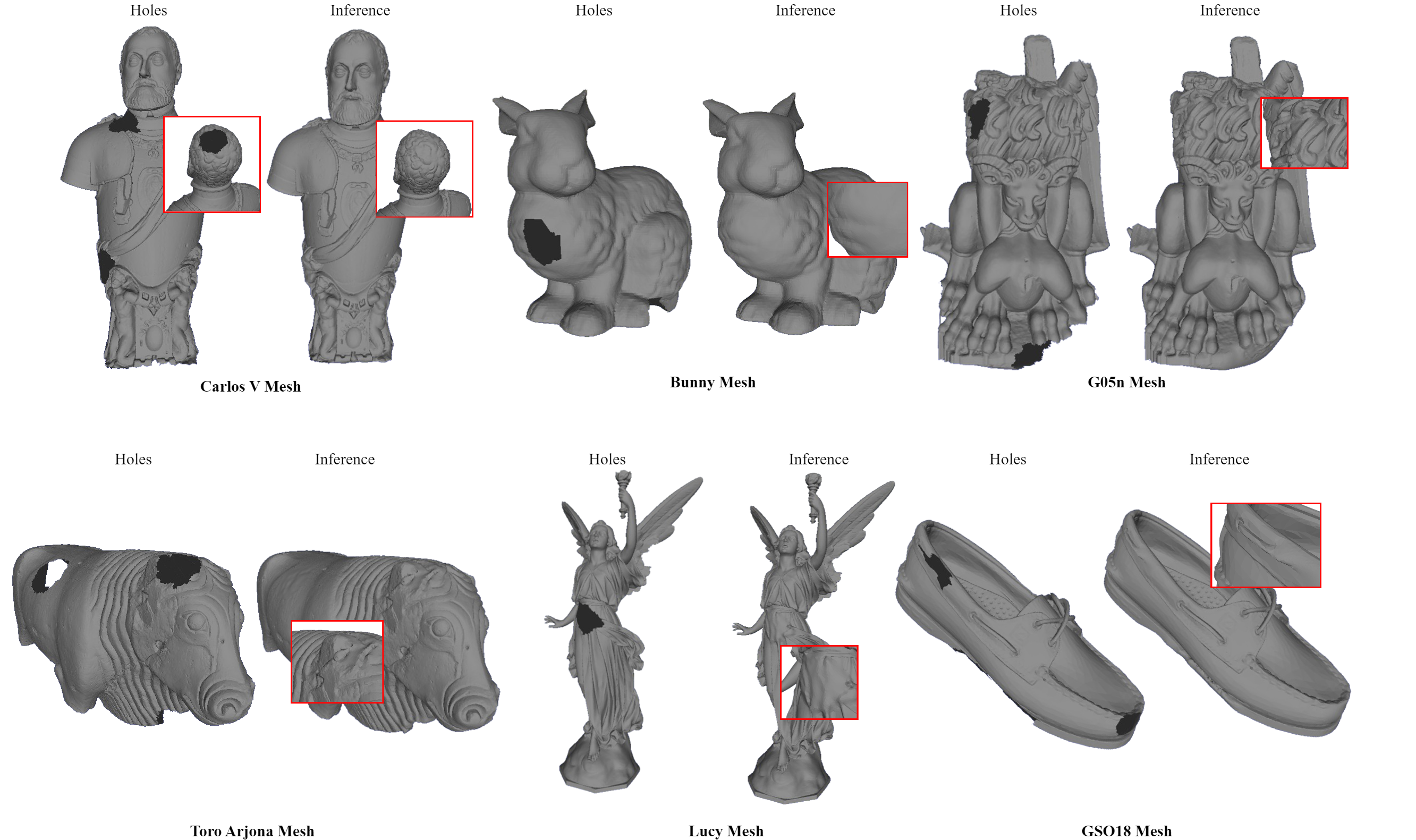}
    \caption{\small Inference results for several models in the dataset.}
    
    \label{fig:qual-examples}
\end{figure*}

Additional visual comparisons of selected meshes from the 3D models dataset are provided in Fig. \ref{fig:qual-examples}. Our method stands out in its ability to repair surfaces with incredibly lifelike and natural results, resulting in meshes that are noticeably smoother than those produced by other cutting-edge methods.

\subsection{Quantitative Results}\label{subsec:quantitative}

In our comprehensive assessment of the method, we concentrated on the comparison of the Hausdorff distance error between the original mesh and the repaired mesh. Table \ref{fig:tabla} contains the summarized results of this comparison of the four models that were previously employed.  These metrics will further facilitate the understanding of the visual reparation results that were previously provided.

Our method consistently demonstrates a significantly reduced maximum distance in comparison to alternative approaches when the Hausdorff distance is taken into account. This finding implies that our method is proficient in mitigating substantial errors within the reconstructed meshes. In addition, the mean distance metric consistently outperforms other methods, suggesting that our reconstructions produce more consistent and stable results across the test models.

The effectiveness of our method in reducing significant errors (as indicated by the smaller maximum distance) while maintaining overall stability and accuracy (as indicated by the smaller mean distance) across a variety of test scenarios is underscored by these results. 

We conducted a comprehensive evaluation of the repairing algorithms on the entire mesh dataset to verify that our method substantially outperforms others in terms of Hausdorff distance. The experiment conditions outlined in Section \ref{subsec:conditions} were taken into account during the execution of these operations.

The mean and maximal Hausdorff distance calculations for a dataset of 959 models are presented in Table \ref{fig:tabla_media}. It is important to note that the SAL method produces \textit{inf}  values as a result of its incapacity to align the inference with the ground truth mesh, which leads to exponentially increased distance values. This problem renders it impossible to automate the measure calculations, which results in the necessity of manual mesh alignment. Furthermore, the Ramesh method has been omitted from this calculation due to the fact that the associated software is supplied as an executable and the code is unavailable. Consequently, it is impossible to script the reparation process across the entire 3D models dataset.

The Lama and Co-mod-GAN inpainting methods exhibit substantial superiority over other state-of-the-art methods, as evidenced by their mean values. This suggests that our method is capable of recovering surfaces with reduced distance differences between their inference and ground truth surfaces. Furthermore, it is important to observe that Co-mod-GAN outperforms Lama, a result that was anticipated by the image loss metrics calculated in Section \ref{section:training}.

\section{Conclusions and future works}\label{section:conclusions}

Based on the results presented in the previous section, SR-CurvANN excels in mesh repair tasks for a variety of reasons. Firstly, by utilizing deep learning techniques, our method is able to acquire intricate surface curvature color patterns, leading to repairs that are highly accurate. Additionally, our thorough repair process focuses on preserving the original mesh shape and details. This results in repaired surfaces that closely resemble the ground truth meshes, without any over-smoothing or unnatural deformations that are often found in other methods. Our quantitative evaluations provide evidence of this. Our method brings unique strengths that greatly contribute to the field of shape completion tasks.

Although SR-CurvANN has shown significant strengths in mesh repair tasks, it is crucial to recognize certain limitations. The effectiveness can be affected by the quality and diversity of the training dataset due to the use of deep learning techniques and extensive training data. Inadequate or skewed training data may lead to less than optimal repairs. In addition, handling severely damaged or incomplete meshes can pose challenges when trying to preserve mesh details during the repair process. A noteworthy example is when meshes show the existence of separate elements or faces in the geometry. We might encounter difficulties in effectively repairing such meshes, as the segmentation and curvature calculations involved necessitate watertight geometries.

These particular instances highlight the importance of further research and development. In addition, the computational resources needed for training and inference in deep learning-based methods, including ours, can be significant, which can restrict the scalability of our approach for real-time or resource-limited applications. Considering future needs, it is important to work on these networks to optimize time and resource utilization.

\begin{table*}[!h]
\centering
\caption{\small Distance measures on different meshes and approaches.}\label{fig:tabla}
\begin{tabular}{ |p{3cm}|p{1.5cm}|p{1.5cm}| }
 \hline
 \multicolumn{3}{|c|}{Face mesh} \\
 \hline
  &max↓&mean↓\\
 \hline
 SAL    & 0.251913 & 0.058967\\
 Ramesh & 0.013503 & \textbf{0.000024}\\
 SPSR   & 0.008064 & 0.000082\\ 
 MeshFix& 0.010498 & 0.000021\\
 Point2Mesh& 0.204235 & 0.012172 \\
 Ours Lama& \textbf{0.008001} & 0.00004\\
 Ours Co-mod-GAN& 0.012252 & 0.000048\\
 \hline
 \multicolumn{3}{|c|}{Christmas Bear mesh} \\
 \hline
  &max↓&mean↓\\
 \hline
 SAL    & 0.422873& 0.087343\\
 Ramesh & 0.049737 & 0.000842\\
 SPSR   & 0.028042 & 0.000974\\
 MeshFix& 0.031404 & 0.000948\\
 Point2Mesh&0.026631 & 0.048666\\
 Ours Lama& \textbf{0.021658} & \textbf{0.000687}\\
 Ours Co-mod-GAN& 0.022175& 0.000687\\
 \hline
\end{tabular}
\quad
\begin{tabular}{ |p{3cm}|p{1.5cm}|p{1.5cm}| }
 \hline
 \multicolumn{3}{|c|}{Armadillo mesh} \\
 \hline
  &max↓&mean↓\\
 \hline
 SAL    & 0.67784 & 0.086474\\
 Ramesh &  0.017833 & 0.000716\\
 SPSR   &  0.057854 & 0.000824\\
 MeshFix& 0.050517 & 0.000726\\
 Point2Mesh& 0.130366 & 0.027310\\
 Ours Lama& \textbf{0.015835} & \textbf{0.000667}\\
 Ours Co-mod-GAN& 0.015835& 0.000668\\
 \hline
 \multicolumn{3}{|c|}{Inmaculada mesh} \\
 \hline
  &max↓&mean↓\\
 \hline
 SAL    & 0.726830 & 0.631150\\
 Ramesh & 0.013539 & 0.000142 \\
 SPSR   & 0.00898 & 0.000532\\
 MeshFix& 0.017925 & 0.000289\\
 Point2Mesh& 0.079542  & 0.030736 \\
 Ours Lama& \textbf{0.013034} & \textbf{0.000132}\\
 Ours Co-mod-GAN& 0.016194 & 0.000138\\
 \hline
\end{tabular}

\end{table*}

\begin{table*}[h!]
\centering
\caption{\small Distance values of the whole mesh dataset (959 models)}\label{fig:tabla_media}
\begin{tabular}{ |p{3cm}|p{3cm}|p{3cm}| }
 \hline
 \multicolumn{3}{|c|}{Hausdorff Distance on dataset (959 models)} \\
 \hline
  &max↓&mean↓\\
 \hline
 SAL    & inf&inf \\
 SPSR   & 0.0886565331506569 & 0.0030312941283506\\ 
 Meshfix & 0.0391596456918267 & 0.0040736090085222 \\
 Point2Mesh& 0.123259509807769 & 0.0097308981768778\\ 
 Ours Lama& 0.035243276105717 & 0.0019252174288026\\
 Ours Co-mod-GAN& \textbf{0,0285419550694646} & \textbf{0,0016561108713763}\\
 \hline
\end{tabular}
\end{table*}

%PLANTILLA---------------------------------------------- 

\section{Acknowledgements}
This work was supported by the Spanish Ministry of Science and Technology under projects PID2023-150070NB-I00 and TED2021-132702B-C21 financed by MCIN/AEI/10.13039/501100011033 and European Regional Development Fund (ERDF).

\clearpage
\bibliographystyle{abbrv}
\bibliography{mainTemplatePDF}

\begin{thebibliography}{10}

\bibitem{attene2010lightweight}
M.~Attene.
\newblock A lightweight approach to repairing digitized polygon meshes.
\newblock {\em The visual computer}, 26:1393--1406, 2010.

\bibitem{atzmon2020sal}
M.~Atzmon and Y.~Lipman.
\newblock Sal: Sign agnostic learning of shapes from raw data.
\newblock In {\em Proceedings of the IEEE/CVF Conference on Computer Vision and Pattern Recognition}, pages 2565--2574, 2020.

\bibitem{bertalmio2000image}
M.~Bertalmio, G.~Sapiro, V.~Caselles, and C.~Ballester.
\newblock Image inpainting.
\newblock In {\em Proceedings of the 27th annual conference on Computer graphics and interactive techniques}, pages 417--424, 2000.

\bibitem{bertalmio2003simultaneous}
M.~Bertalmio, L.~Vese, G.~Sapiro, and S.~Osher.
\newblock Simultaneous structure and texture image inpainting.
\newblock {\em IEEE transactions on image processing}, 12(8):882--889, 2003.

\bibitem{botsch2004remeshing}
M.~Botsch and L.~Kobbelt.
\newblock A remeshing approach to multiresolution modeling.
\newblock In {\em Proceedings of the 2004 Eurographics/ACM SIGGRAPH symposium on Geometry processing}, pages 185--192, 2004.

\bibitem{Botsch2010polygon}
M.~Botsch, L.~Kobbelt, M.~Pauly, P.~Alliez, and B.~L{\'e}vy.
\newblock {\em Polygon mesh processing}.
\newblock CRC press, 2010.

\bibitem{brunton2009filling}
A.~Brunton, S.~Wuhrer, C.~Shu, P.~Bose, and E.~D. Demaine.
\newblock Filling holes in triangular meshes by curve unfolding.
\newblock In {\em 2009 IEEE International Conference on Shape Modeling and Applications}, pages 66--72. IEEE, 2009.

\bibitem{bugeau2010comprehensive}
A.~Bugeau, M.~Bertalm{\'\i}o, V.~Caselles, and G.~Sapiro.
\newblock A comprehensive framework for image inpainting.
\newblock {\em IEEE transactions on image processing}, 19(10):2634--2645, 2010.

\bibitem{centin2015rameshcleaner}
M.~Centin and A.~Signoroni.
\newblock Rameshcleaner: conservative fixing of triangular meshes.
\newblock 2015.

\bibitem{chang2015shapenet}
A.~X. Chang, T.~Funkhouser, L.~Guibas, P.~Hanrahan, Q.~Huang, Z.~Li, S.~Savarese, M.~Savva, S.~Song, H.~Su, et~al.
\newblock Shapenet: An information-rich 3d model repository.
\newblock {\em arXiv preprint arXiv:1512.03012}, 2015.

\bibitem{chao2010simple}
I.~Chao, U.~Pinkall, P.~Sanan, and P.~Schr{\"o}der.
\newblock A simple geometric model for elastic deformations.
\newblock {\em ACM transactions on graphics (TOG)}, 29(4):1--6, 2010.

\bibitem{chu2021unsupervised}
L.~Chu, H.~Pan, and W.~Wang.
\newblock Unsupervised shape completion via deep prior in the neural tangent kernel perspective.
\newblock {\em ACM Transactions on Graphics (TOG)}, 40(3):1--17, 2021.

\bibitem{collins2022abo}
J.~Collins, S.~Goel, K.~Deng, A.~Luthra, L.~Xu, E.~Gundogdu, X.~Zhang, T.~F.~Y. Vicente, T.~Dideriksen, H.~Arora, et~al.
\newblock Abo: Dataset and benchmarks for real-world 3d object understanding.
\newblock In {\em Proceedings of the IEEE/CVF conference on computer vision and pattern recognition}, pages 21126--21136, 2022.

\bibitem{criminisi2004region}
A.~Criminisi, P.~P{\'e}rez, and K.~Toyama.
\newblock Region filling and object removal by exemplar-based image inpainting.
\newblock {\em IEEE Transactions on image processing}, 13(9):1200--1212, 2004.

\bibitem{dai2017scannet}
A.~Dai, A.~X. Chang, M.~Savva, M.~Halber, T.~Funkhouser, and M.~Nie{\ss}ner.
\newblock Scannet: Richly-annotated 3d reconstructions of indoor scenes.
\newblock In {\em Proceedings of the IEEE conference on computer vision and pattern recognition}, pages 5828--5839, 2017.

\bibitem{dai2017shape}
A.~Dai, C.~Ruizhongtai~Qi, and M.~Nie{\ss}ner.
\newblock Shape completion using 3d-encoder-predictor cnns and shape synthesis.
\newblock In {\em Proceedings of the IEEE conference on computer vision and pattern recognition}, pages 5868--5877, 2017.

\bibitem{davis2002filling}
J.~Davis, S.~R. Marschner, M.~Garr, and M.~Levoy.
\newblock Filling holes in complex surfaces using volumetric diffusion.
\newblock In {\em Proceedings. First international symposium on 3d data processing visualization and transmission}, pages 428--441. IEEE, 2002.

\bibitem{deitke2023objaverse}
M.~Deitke, D.~Schwenk, J.~Salvador, L.~Weihs, O.~Michel, E.~VanderBilt, L.~Schmidt, K.~Ehsani, A.~Kembhavi, and A.~Farhadi.
\newblock Objaverse: A universe of annotated 3d objects.
\newblock In {\em Proceedings of the IEEE/CVF Conference on Computer Vision and Pattern Recognition}, pages 13142--13153, 2023.

\bibitem{ding2018perceptually}
D.~Ding, S.~Ram, and J.~J. Rodriguez.
\newblock Perceptually aware image inpainting.
\newblock {\em Pattern Recognition}, 83:174--184, 2018.

\bibitem{downs2022google}
L.~Downs, A.~Francis, N.~Koenig, B.~Kinman, R.~Hickman, K.~Reymann, T.~B. McHugh, and V.~Vanhoucke.
\newblock Google scanned objects: A high-quality dataset of 3d scanned household items.
\newblock In {\em 2022 International Conference on Robotics and Automation (ICRA)}, pages 2553--2560. IEEE, 2022.

\bibitem{efros2001image}
A.~A. Efros and W.~T. Freeman.
\newblock Image quilting for texture synthesis and transfer.
\newblock In {\em Proceedings of the 28th annual conference on Computer graphics and interactive techniques}, pages 341--346, 2001.

\bibitem{fan2017point}
H.~Fan, H.~Su, and L.~J. Guibas.
\newblock A point set generation network for 3d object reconstruction from a single image.
\newblock In {\em Proceedings of the IEEE conference on computer vision and pattern recognition}, pages 605--613, 2017.

\bibitem{fang1995delaunay}
T.-P. Fang and L.~A. Piegl.
\newblock Delaunay triangulation in three dimensions.
\newblock {\em IEEE Computer Graphics and Applications}, 15(5):62--69, 1995.

\bibitem{geiger2012we}
A.~Geiger, P.~Lenz, and R.~Urtasun.
\newblock Are we ready for autonomous driving? the kitti vision benchmark suite.
\newblock In {\em 2012 IEEE conference on computer vision and pattern recognition}, pages 3354--3361. IEEE, 2012.

\bibitem{gisbert2023inpainting}
G.~Gisbert, R.~Chaine, and D.~Coeurjolly.
\newblock Inpainting holes in folded fabric meshes.
\newblock {\em Computers \& Graphics}, 114:201--209, 2023.

\bibitem{guo2006filling}
T.-Q. Guo, J.-J. Li, J.-G. Weng, and Y.-T. Zhuang.
\newblock Filling holes in complex surfaces using oriented voxel diffusion.
\newblock In {\em 2006 International Conference on Machine Learning and Cybernetics}, pages 4370--4375. IEEE, 2006.

\bibitem{hanocka2020point2mesh}
R.~Hanocka, G.~Metzer, R.~Giryes, and D.~Cohen-Or.
\newblock Point2mesh: A self-prior for deformable meshes.
\newblock {\em arXiv preprint arXiv:2005.11084}, 2020.

\bibitem{harary2014context}
G.~Harary, A.~Tal, and E.~Grinspun.
\newblock Context-based coherent surface completion.
\newblock {\em ACM Transactions on Graphics (TOG)}, 33(1):1--12, 2014.

\bibitem{hattori2022learning}
S.~Hattori, T.~Yatagawa, Y.~Ohtake, and H.~Suzuki.
\newblock Learning self-prior for mesh denoising using dual graph convolutional networks.
\newblock In {\em European Conference on Computer Vision}, pages 363--379. Springer, 2022.

\bibitem{hattori2024learning}
S.~Hattori, T.~Yatagawa, Y.~Ohtake, and H.~Suzuki.
\newblock Learning self-prior for mesh inpainting using self-supervised graph convolutional networks.
\newblock {\em IEEE Transactions on Visualization and Computer Graphics}, 2024.

\bibitem{hermoza20183d}
R.~Hermoza and I.~Sipiran.
\newblock 3d reconstruction of incomplete archaeological objects using a generative adversarial network.
\newblock In {\em Proceedings of Computer Graphics International 2018}, pages 5--11. 2018.

\bibitem{hernandez2023deep}
M.~Hern{\'a}ndez-Bautista and F.~Melero.
\newblock Deep learning of curvature features for shape completion.
\newblock {\em Computers \& Graphics}, 2023.

\bibitem{hore2010image}
A.~Hore and D.~Ziou.
\newblock Image quality metrics: Psnr vs. ssim.
\newblock In {\em 2010 20th international conference on pattern recognition}, pages 2366--2369. IEEE, 2010.

\bibitem{hu20203d}
T.~Hu, Z.~Han, and M.~Zwicker.
\newblock 3d shape completion with multi-view consistent inference.
\newblock In {\em Proceedings of the AAAI conference on artificial intelligence}, volume~34, pages 10997--11004, 2020.

\bibitem{ji2020image}
J.~Ji, B.~Zhong, and K.-K. Ma.
\newblock Image interpolation using multi-scale attention-aware inception network.
\newblock {\em IEEE Transactions on Image Processing}, 29:9413--9428, 2020.

\bibitem{jiang2021transgan}
Y.~Jiang, S.~Chang, and Z.~Wang.
\newblock Transgan: Two transformers can make one strong gan.
\newblock {\em arXiv preprint arXiv:2102.07074}, 1(3), 2021.

\bibitem{kazhdan2013screened}
M.~Kazhdan and H.~Hoppe.
\newblock Screened poisson surface reconstruction.
\newblock {\em ACM Transactions on Graphics (ToG)}, 32(3):1--13, 2013.

\bibitem{koch2019abc}
S.~Koch, A.~Matveev, Z.~Jiang, F.~Williams, A.~Artemov, E.~Burnaev, M.~Alexa, D.~Zorin, and D.~Panozzo.
\newblock Abc: A big cad model dataset for geometric deep learning.
\newblock In {\em Proceedings of the IEEE/CVF conference on computer vision and pattern recognition}, pages 9601--9611, 2019.

\bibitem{li2022mat}
W.~Li, Z.~Lin, K.~Zhou, L.~Qi, Y.~Wang, and J.~Jia.
\newblock Mat: Mask-aware transformer for large hole image inpainting.
\newblock In {\em Proceedings of the IEEE/CVF conference on computer vision and pattern recognition}, pages 10758--10768, 2022.

\bibitem{liang2001real}
L.~Liang, C.~Liu, Y.-Q. Xu, B.~Guo, and H.-Y. Shum.
\newblock Real-time texture synthesis by patch-based sampling.
\newblock {\em ACM Transactions on Graphics (ToG)}, 20(3):127--150, 2001.

\bibitem{liepa2003filling}
P.~Liepa.
\newblock Filling holes in meshes.
\newblock In {\em Proceedings of the 2003 Eurographics/ACM SIGGRAPH symposium on Geometry processing}, pages 200--205, 2003.

\bibitem{liu2018image}
G.~Liu, F.~A. Reda, K.~J. Shih, T.-C. Wang, A.~Tao, and B.~Catanzaro.
\newblock Image inpainting for irregular holes using partial convolutions.
\newblock In {\em Proceedings of the European conference on computer vision (ECCV)}, pages 85--100, 2018.

\bibitem{long2015fully}
J.~Long, E.~Shelhamer, and T.~Darrell.
\newblock Fully convolutional networks for semantic segmentation.
\newblock In {\em Proceedings of the IEEE conference on computer vision and pattern recognition}, pages 3431--3440, 2015.

\bibitem{maggiordomo2023texture}
A.~Maggiordomo, P.~Cignoni, and M.~Tarini.
\newblock Texture inpainting for photogrammetric models.
\newblock In {\em Computer Graphics Forum}. Wiley Online Library, 2023.

\bibitem{mao2016image}
X.~Mao, C.~Shen, and Y.-B. Yang.
\newblock Image restoration using very deep convolutional encoder-decoder networks with symmetric skip connections.
\newblock {\em Advances in neural information processing systems}, 29, 2016.

\bibitem{mirzaei2023reference}
A.~Mirzaei, T.~Aumentado-Armstrong, M.~A. Brubaker, J.~Kelly, A.~Levinshtein, K.~G. Derpanis, and I.~Gilitschenski.
\newblock Reference-guided controllable inpainting of neural radiance fields.
\newblock In {\em Proceedings of the IEEE/CVF International Conference on Computer Vision}, pages 17815--17825, 2023.

\bibitem{park2019deepsdf}
J.~J. Park, P.~Florence, J.~Straub, R.~Newcombe, and S.~Lovegrove.
\newblock Deepsdf: Learning continuous signed distance functions for shape representation.
\newblock In {\em Proceedings of the IEEE/CVF conference on computer vision and pattern recognition}, pages 165--174, 2019.

\bibitem{park2006surface}
S.~Park, X.~Guo, H.~Shin, and H.~Qin.
\newblock Surface completion for shape and appearance.
\newblock {\em The Visual Computer}, 22:168--180, 2006.

\bibitem{pathak2016context}
D.~Pathak, P.~Krahenbuhl, J.~Donahue, T.~Darrell, and A.~A. Efros.
\newblock Context encoders: Feature learning by inpainting.
\newblock In {\em Proceedings of the IEEE conference on computer vision and pattern recognition}, pages 2536--2544, 2016.

\bibitem{perez2021repairing}
E.~P{\'e}rez, S.~Salamanca, P.~Merch{\'a}n, and A.~Ad{\'a}n.
\newblock Repairing 3d models obtained from range sensors.
\newblock {\em IEEE Access}, 9:43474--43493, 2021.

\bibitem{poole2022dreamfusion}
B.~Poole, A.~Jain, J.~T. Barron, and B.~Mildenhall.
\newblock Dreamfusion: Text-to-3d using 2d diffusion.
\newblock {\em arXiv preprint arXiv:2209.14988}, 2022.

\bibitem{qiang2010hole}
H.~Qiang, Z.~Shusheng, B.~Xiaoliang, and Z.~Xin.
\newblock Hole filling based on local surface approximation.
\newblock In {\em 2010 International Conference on Computer Application and System Modeling (ICCASM 2010)}, volume~3, pages V3--242. IEEE, 2010.

\bibitem{qin2020face}
J.~Qin, H.~Bai, and Y.~Zhao.
\newblock Face inpainting network for large missing regions based on weighted facial similarity.
\newblock {\em Neurocomputing}, 386:54--62, 2020.

\bibitem{radford2021learning}
A.~Radford, J.~W. Kim, C.~Hallacy, A.~Ramesh, G.~Goh, S.~Agarwal, G.~Sastry, A.~Askell, P.~Mishkin, J.~Clark, et~al.
\newblock Learning transferable visual models from natural language supervision.
\newblock In {\em International conference on machine learning}, pages 8748--8763. PMLR, 2021.

\bibitem{ramesh2022hierarchical}
A.~Ramesh, P.~Dhariwal, A.~Nichol, C.~Chu, and M.~Chen.
\newblock Hierarchical text-conditional image generation with clip latents.
\newblock {\em arXiv preprint arXiv:2204.06125}, 2022.

\bibitem{riegler2017octnet}
G.~Riegler, A.~Osman~Ulusoy, and A.~Geiger.
\newblock Octnet: Learning deep 3d representations at high resolutions.
\newblock In {\em Proceedings of the IEEE conference on computer vision and pattern recognition}, pages 3577--3586, 2017.

\bibitem{rombach2021highresolution}
R.~Rombach, A.~Blattmann, D.~Lorenz, P.~Esser, and B.~Ommer.
\newblock High-resolution image synthesis with latent diffusion models, 2021.

\bibitem{ronneberger2015u}
O.~Ronneberger, P.~Fischer, and T.~Brox.
\newblock U-net: Convolutional networks for biomedical image segmentation.
\newblock In {\em Medical image computing and computer-assisted intervention--MICCAI 2015: 18th international conference, Munich, Germany, October 5-9, 2015, proceedings, part III 18}, pages 234--241. Springer, 2015.

\bibitem{sarmad2019rl}
M.~Sarmad, H.~J. Lee, and Y.~M. Kim.
\newblock Rl-gan-net: A reinforcement learning agent controlled gan network for real-time point cloud shape completion.
\newblock In {\em Proceedings of the IEEE/CVF Conference on Computer Vision and Pattern Recognition}, pages 5898--5907, 2019.

\bibitem{sasaki2017joint}
K.~Sasaki, S.~Iizuka, E.~Simo-Serra, and H.~Ishikawa.
\newblock Joint gap detection and inpainting of line drawings.
\newblock In {\em Proceedings of the IEEE conference on computer vision and pattern recognition}, pages 5725--5733, 2017.

\bibitem{shapira2008consistent}
L.~Shapira, A.~Shamir, and D.~Cohen-Or.
\newblock Consistent mesh partitioning and skeletonisation using the shape diameter function.
\newblock {\em The Visual Computer}, 24:249--259, 2008.

\bibitem{sharf2004context}
A.~Sharf, M.~Alexa, and D.~Cohen-Or.
\newblock Context-based surface completion.
\newblock In {\em ACM SIGGRAPH 2004 Papers}, pages 878--887. 2004.

\bibitem{sipiran2022data}
I.~Sipiran, A.~Mendoza, A.~Apaza, and C.~Lopez.
\newblock Data-driven restoration of digital archaeological pottery with point cloud analysis.
\newblock {\em International Journal of Computer Vision}, 130(9):2149--2165, 2022.

\bibitem{stutz2018learning}
D.~Stutz and A.~Geiger.
\newblock Learning 3d shape completion from laser scan data with weak supervision.
\newblock In {\em Proceedings of the IEEE Conference on Computer Vision and Pattern Recognition}, pages 1955--1964, 2018.

\bibitem{suvorov2021resolution}
R.~Suvorov, E.~Logacheva, A.~Mashikhin, A.~Remizova, A.~Ashukha, A.~Silvestrov, N.~Kong, H.~Goka, K.~Park, and V.~Lempitsky.
\newblock Resolution-robust large mask inpainting with fourier convolutions.
\newblock {\em arXiv preprint arXiv:2109.07161}, 2021.

\bibitem{vichitvejpaisal2014surface}
P.~Vichitvejpaisal and P.~Kanongchaiyos.
\newblock Surface completion using laplacian transform.
\newblock {\em Engineering Journal}, 18(1):129--144, 2014.

\bibitem{wang20203d}
X.~Wang, D.~Xu, and F.~Gu.
\newblock 3d model inpainting based on 3d deep convolutional generative adversarial network.
\newblock {\em IEEE Access}, 8:170355--170363, 2020.

\bibitem{weber2024}
E.~Weber, A.~Holynski, V.~Jampani, S.~Saxena, N.~Snavely, A.~Kar, and A.~Kanazawa.
\newblock Nerfiller: Completing scenes via generative 3d inpainting.
\newblock In {\em Proceedings of the IEEE/CVF Conference on Computer Vision and Pattern Recognition (CVPR)}, pages 20731--20741, June 2024.

\bibitem{weber2024nerfiller}
E.~Weber, A.~Holynski, V.~Jampani, S.~Saxena, N.~Snavely, A.~Kar, and A.~Kanazawa.
\newblock Nerfiller: Completing scenes via generative 3d inpainting.
\newblock In {\em Proceedings of the IEEE/CVF Conference on Computer Vision and Pattern Recognition}, pages 20731--20741, 2024.

\bibitem{wei2010integrated}
M.~Wei, J.~Wu, and M.~Pang.
\newblock An integrated approach to filling holes in meshes.
\newblock In {\em 2010 International Conference on Artificial Intelligence and Computational Intelligence}, volume~3, pages 306--310. IEEE, 2010.

\bibitem{wen2020point}
X.~Wen, T.~Li, Z.~Han, and Y.-S. Liu.
\newblock Point cloud completion by skip-attention network with hierarchical folding.
\newblock In {\em Proceedings of the IEEE/CVF conference on computer vision and pattern recognition}, pages 1939--1948, 2020.

\bibitem{williams2019deep}
F.~Williams, T.~Schneider, C.~Silva, D.~Zorin, J.~Bruna, and D.~Panozzo.
\newblock Deep geometric prior for surface reconstruction.
\newblock In {\em Proceedings of the IEEE/CVF Conference on Computer Vision and Pattern Recognition}, pages 10130--10139, 2019.

\bibitem{wu2008automatic}
X.~J. Wu, M.~Y. Wang, and B.~Han.
\newblock An automatic hole-filling algorithm for polygon meshes.
\newblock {\em Computer-Aided Design and Applications}, 5(6):889--899, 2008.

\bibitem{wu20153d}
Z.~Wu, S.~Song, A.~Khosla, F.~Yu, L.~Zhang, X.~Tang, and J.~Xiao.
\newblock 3d shapenets: A deep representation for volumetric shapes.
\newblock In {\em Proceedings of the IEEE conference on computer vision and pattern recognition}, pages 1912--1920, 2015.

\bibitem{xiao2023}
A.~Xiao, J.~Huang, D.~Guan, X.~Zhang, S.~Lu, and L.~Shao.
\newblock Unsupervised point cloud representation learning with deep neural networks: A survey.
\newblock {\em IEEE Trans. Pattern Anal. Mach. Intell.}, 45(9):11321–11339, sep 2023.

\bibitem{xie2019image}
C.~Xie, S.~Liu, C.~Li, M.-M. Cheng, W.~Zuo, X.~Liu, S.~Wen, and E.~Ding.
\newblock Image inpainting with learnable bidirectional attention maps.
\newblock In {\em Proceedings of the IEEE/CVF international conference on computer vision}, pages 8858--8867, 2019.

\bibitem{yang2018foldingnet}
Y.~Yang, C.~Feng, Y.~Shen, and D.~Tian.
\newblock Foldingnet: Point cloud auto-encoder via deep grid deformation.
\newblock In {\em Proceedings of the IEEE conference on computer vision and pattern recognition}, pages 206--215, 2018.

\bibitem{yu2018generative}
J.~Yu, Z.~Lin, J.~Yang, X.~Shen, X.~Lu, and T.~S. Huang.
\newblock Generative image inpainting with contextual attention.
\newblock In {\em Proceedings of the IEEE conference on computer vision and pattern recognition}, pages 5505--5514, 2018.

\bibitem{yu2021frechet}
Y.~Yu, W.~Zhang, and Y.~Deng.
\newblock Frechet inception distance (fid) for evaluating gans.
\newblock {\em China University of Mining Technology Beijing Graduate School}, 2021.

\bibitem{yuan2018pcn}
W.~Yuan, T.~Khot, D.~Held, C.~Mertz, and M.~Hebert.
\newblock Pcn: Point completion network.
\newblock In {\em 2018 international conference on 3D vision (3DV)}, pages 728--737. IEEE, 2018.

\bibitem{zeng2019learning}
Y.~Zeng, J.~Fu, H.~Chao, and B.~Guo.
\newblock Learning pyramid-context encoder network for high-quality image inpainting.
\newblock In {\em Proceedings of the IEEE/CVF conference on computer vision and pattern recognition}, pages 1486--1494, 2019.

\bibitem{zeng2022aggregated}
Y.~Zeng, J.~Fu, H.~Chao, and B.~Guo.
\newblock Aggregated contextual transformations for high-resolution image inpainting.
\newblock {\em IEEE Transactions on Visualization and Computer Graphics}, 2022.

\bibitem{zeng2021cr}
Y.~Zeng, Z.~Lin, H.~Lu, and V.~M. Patel.
\newblock Cr-fill: Generative image inpainting with auxiliary contextual reconstructionyu2018generative.
\newblock In {\em Proceedings of the IEEE/CVF international conference on computer vision}, pages 14164--14173, 2021.

\bibitem{zhang2021unsupervised}
J.~Zhang, X.~Chen, Z.~Cai, L.~Pan, H.~Zhao, S.~Yi, C.~K. Yeo, B.~Dai, and C.~C. Loy.
\newblock Unsupervised 3d shape completion through gan inversion.
\newblock In {\em Proceedings of the IEEE/CVF Conference on Computer Vision and Pattern Recognition}, pages 1768--1777, 2021.

\bibitem{zhang2018unreasonable}
R.~Zhang, P.~Isola, A.~A. Efros, E.~Shechtman, and O.~Wang.
\newblock The unreasonable effectiveness of deep features as a perceptual metric.
\newblock In {\em Proceedings of the IEEE conference on computer vision and pattern recognition}, pages 586--595, 2018.

\bibitem{zhang2017demeshnet}
S.~Zhang, R.~He, Z.~Sun, and T.~Tan.
\newblock Demeshnet: Blind face inpainting for deep meshface verification.
\newblock {\em IEEE Transactions on Information Forensics and Security}, 13(3):637--647, 2017.

\bibitem{zhao2021large}
S.~Zhao, J.~Cui, Y.~Sheng, Y.~Dong, X.~Liang, E.~I. Chang, and Y.~Xu.
\newblock Large scale image completion via co-modulated generative adversarial networks.
\newblock {\em arXiv preprint arXiv:2103.10428}, 2021.

\bibitem{zheng2022bridging}
C.~Zheng, T.-J. Cham, J.~Cai, and D.~Phung.
\newblock Bridging global context interactions for high-fidelity image completion.
\newblock In {\em Proceedings of the IEEE/CVF Conference on Computer Vision and Pattern Recognition}, pages 11512--11522, 2022.

\bibitem{zhou2017places}
B.~Zhou, A.~Lapedriza, A.~Khosla, A.~Oliva, and A.~Torralba.
\newblock Places: A 10 million image database for scene recognition.
\newblock {\em IEEE transactions on pattern analysis and machine intelligence}, 40(6):1452--1464, 2017.

\bibitem{zhu2021image}
M.~Zhu, D.~He, X.~Li, C.~Li, F.~Li, X.~Liu, E.~Ding, and Z.~Zhang.
\newblock Image inpainting by end-to-end cascaded refinement with mask awareness.
\newblock {\em IEEE Transactions on Image Processing}, 30:4855--4866, 2021.

\end{thebibliography}
\end{document}